\documentclass[a4paper]{jpconf}
\usepackage{graphicx}
\usepackage[figuresright]{rotating}
\def\beq{\begin{equation}}
\def\eeq{\end{equation}}
\def\ga{\mathrel{\raise.3ex\hbox{$>$\kern-.75em\lower1ex\hbox{$\sim$}}}}
\def\la{\mathrel{\raise.3ex\hbox{$<$\kern-.75em\lower1ex\hbox{$\sim$}}}}
\newcommand{\gmt}{$(g-2)_\mu$}
\newcommand{\bsg}{BR($b \to s \gamma$)}

\newcommand{\bmm}{BR($B_s \to \mu^+\mu^-$)}
\newcommand{\ssi}{\sigma^{\rm SI}_p}
\newcommand{\Och}{\ensuremath{\Omega_\chi h^2}}

\newcommand{\Mh}{M_h}

\newcommand{\gev}{\,\, \mathrm{GeV}}
\newcommand{\tev}{\,\, \mathrm{TeV}}
\newcommand{\tb}{\tan\beta}

\newcommand{\mneu}[1]{m_{\tilde \chi^0_{#1}}}
\newcommand{\ETslash}{/ \hspace{-.7em} E_T}

\begin{document}
\title{The impact of XENON100 and the LHC on Supersymmetric Dark Matter$^*$}

\author{Keith A. Olive}

\address{William\,I.\,Fine\,Theoretical\,Physics\,Institute,\\ School of Physics\,and\,Astronomy,\,University\,of\,Minnesota,\\Minneapolis,\,Minnesota\,55455,\,USA\\
 $^*$To be published in the Proceedings of the 7th DSU Conference, Beijing China}

\ead{olive@umn.edu}

\begin{abstract}
\vskip - 2.7in
\rightline{UMN--TH--3033/12}
\rightline{FTPI--MINN--12/06}
\vskip +2.5in
The effect of 2010 and  2011 LHC data are discussed in connection to the potential for the direct detection of supersymmetric dark matter.
The impact of the recent XENON100 results
are contrasted to these predictions.
\end{abstract}

\section{Introduction}
The minimal supersymmetric Standard Model (MSSM) 
has over 100 undetermined parameters, which are
mainly associated with the breaking of supersymmetry. However, it is often assumed that the
soft supersymmetry-breaking parameters have some
universality properties. These may include the universality of gaugino masses, $M_a = m_{1/2}$,
trilinear supersymmetry-breaking mass parameters, $A_f = A_0$, and soft scalar masses. 
$m^2_{ij} = \delta_{ij} m_0^2$.
 The simplified version of the MSSM in which universality in input 
at the grand unification scale is 
called the constrained MSSM (CMSSM) \cite{funnel,cmssm,efgosi,cmssm2,cmssmwmap,mc1,mc2,mc3}.
The sparticle spectrum is run down to the electroweak scale
and radiatively induces electroweak symmetry breaking (EWSB).

Minimization of the Higgs potential leads to two conditions at the weak scale which 
can be expressed as
\beq
\mu^2=\frac{m_1^2-m_2^2\tan^2\beta+\frac{1}{2}m_Z^2(1-\tan^2\beta)+\Delta_{\mu}^{(1)}}{\tan^2\beta-1+\Delta_{\mu}^{(2)}},
\label{eq:mu}
\eeq
and 
\beq
B \mu = (m_1^2+m_2^2+2\mu^2)\sin 2\beta +\Delta_B
\label{eq:muB}
\eeq  
where $\mu$ is the Higgs mixing parameter, $B$ is the associated
supersymmetry-breaking bilinear mass, $\tan \beta$ is the ratio of the two Higgs
vacuum expectation values, and $\Delta_B$ and $\Delta_\mu^{(1,2)}$ are loop
corrections~\cite{Barger:1993gh,deBoer:1994he,Carena:2001fw}.
The combination $B\mu$ can be related to the Higgs pseudo-scalar mass, $m_A$. 
While one can
choose to include $B$ and $\mu$ (or $m_A$ and $\mu$) as free input parameters and
calculate the two Higgs expectation values, or $m_Z$ and $\tan \beta$, it is
more common to use these equations to calculate $\mu$ and $B$ upon assuming
a value of $\tan \beta$ and the measured value of $m_Z$.
Thus upon assuming radiative electroweak symmetry breaking, the CMSSM
is a 4 parameter theory ($m_{1/2}, m_0, A_0$ and $\tan \beta$). In addition the
sign of the $\mu$ term must also be specified.  

An often considered generalization of the CMSSM allows for non-universal Higgs masses (NUHM)
\cite{nuhm2,nuhm1,nuhm1-2}. By allowing, both Higgs soft masses $m_1$ and $m_2$ to differ
from $m_0$ (NUHM2), one can effectively choose both $m_A$ and $\mu$ as free parameters
as is seen from the electroweak conditions given above. One may also choose a subset of these
models and take $m_1 = m_2 \ne m_0$  in which case either $m_A$ or $\mu$ can
be chosen in addition to the four CMSSM parameters (NUHM1). 

It is also possible to consider a very constrained version of the MSSM (VCMSSM) \cite{vcmssm}
by applying the relation $B_0 = A_0 - m_0$ as expected from minimal supergravity \cite{mSUGRA}.
In this case, $\mu$ and $\tan \beta$ are derived from the electroweak vacuum conditions
and the theory has only three free parameters (and the sign of $\mu$). 
True models based on minimal supergravity (mSUGRA) impose in addition 
the relation between the gravitino mass and $m_0$, namely $m_{3/2} = m_0$.
In these models, it is often the gravitino which ends up as the lightest supersymmetric particle
(LSP) and therefore the dark matter candidate.

\begin{figure}
  \includegraphics[width=.5\textwidth]{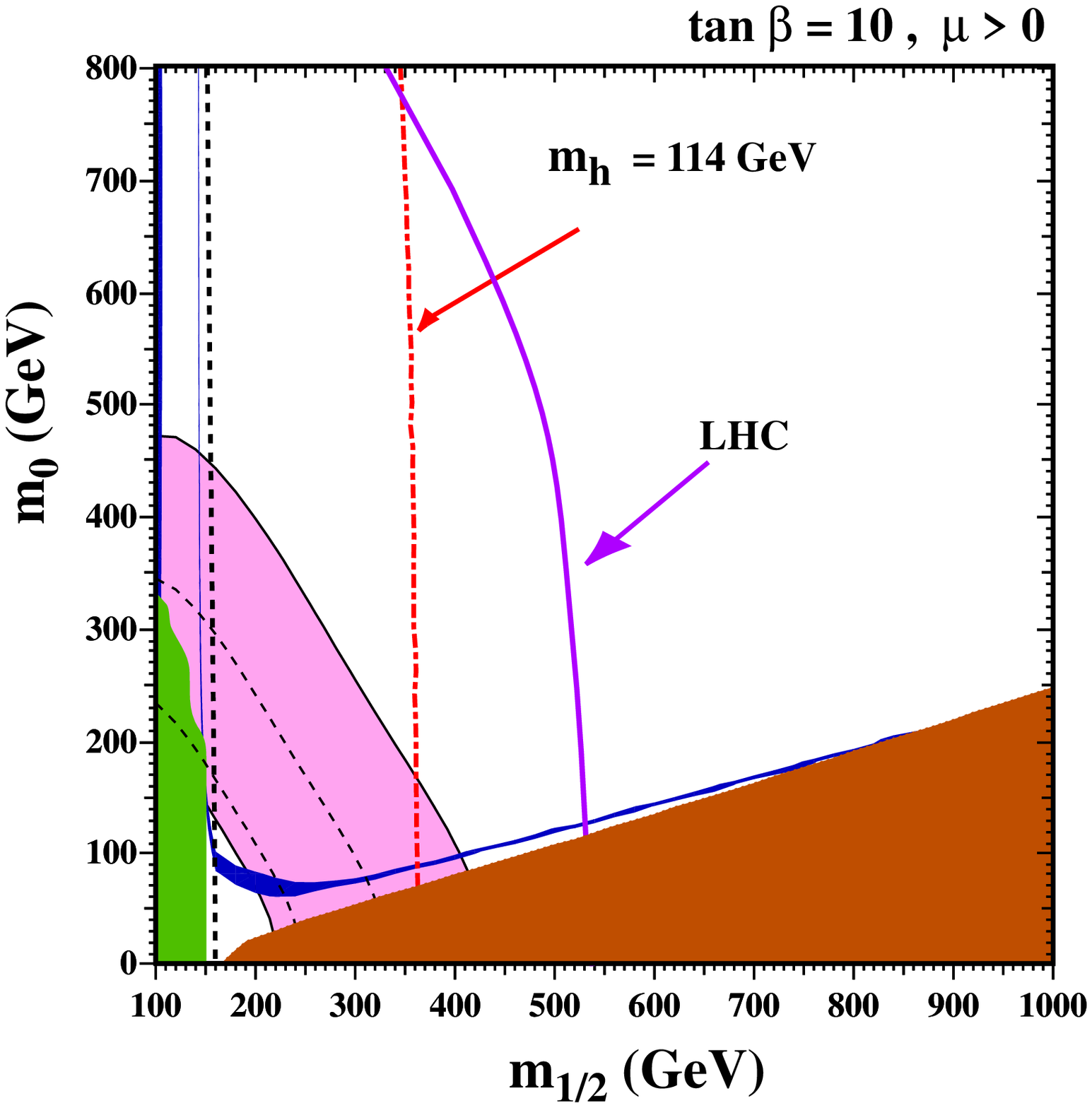}
    \includegraphics[width=.5\textwidth]{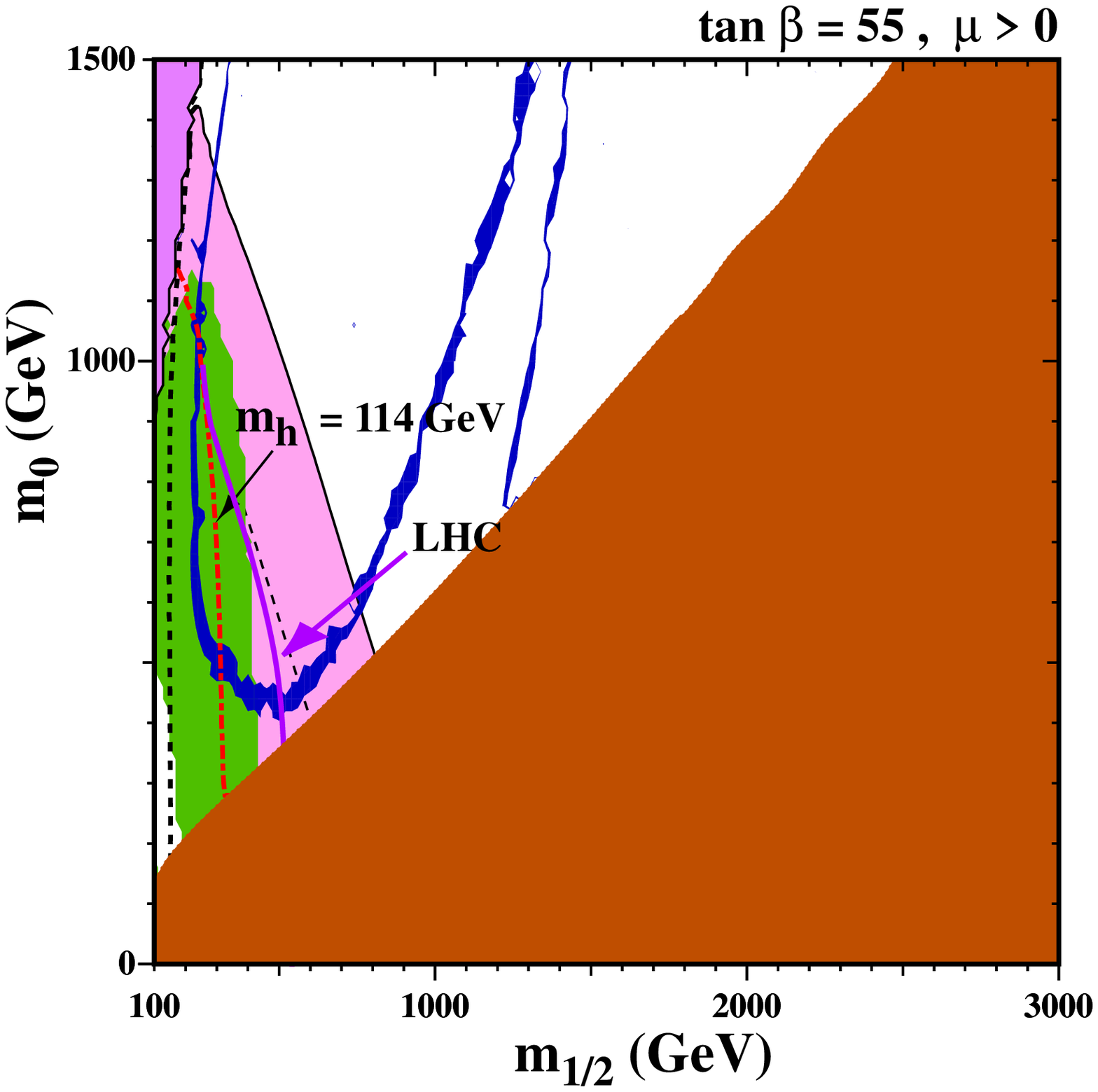}
  \caption{\it The $(m_{1/2}, m_0)$ planes for  (a) $\tan \beta = 10$ and  $\mu > 0$, 
assuming $A_0 = 0, m_t = 173.1$~GeV and
$m_b(m_b)^{\overline {MS}}_{SM} = 4.25$~GeV. The near-vertical (red)
dot-dashed lines are the contours $m_h = 114$~GeV, and the near-vertical (black) dashed
line is the contour $m_{\chi^\pm} = 104$~GeV.  The medium (dark
green) shaded region is excluded by $b \to s
\gamma$, and the dark (blue) shaded area is the cosmologically
preferred region. In the dark
(brick red) shaded region, the LSP is the charged ${\tilde \tau}_1$. The
region allowed by the E821 measurement of $g_\mu -2$ at the 2-$\sigma$
level, is shaded (pink) and bounded by solid black lines, with dashed
lines indicating the 1-$\sigma$ ranges.
The curves marked LHC show the 95\& CL exclusion region (to the left of the curves)
for LHC sparticle searches at 1/fb. In (b), $\tan \beta= 55$. Here, in the upper left corner,
the region with no EWSB is shaded dark pink.}
  \label{fig:UHM}
\end{figure}

For given values of $\tan \beta$, $A_0$,  and $sgn(\mu)$, the regions of the CMSSM
parameter space that yield an
acceptable relic density and satisfy other phenomenological constraints
may be displayed in the  $(m_{1/2}, m_0)$ plane.
In Fig. \ref{fig:UHM}a,  the dark (blue)
shaded region corresponds to that portion of the CMSSM plane
with $\tan \beta = 10$, $A_0 = 0$, and $\mu > 0$ such that the computed
relic density yields the WMAP value \cite{wmap} of 
\beq
\Omega h^2 = 0.111 \pm 0.006 .
\eeq
The bulk region at relatively low values of 
$m_{1/2}$ and $m_0$,  tapers off
as $m_{1/2}$ is increased.  At higher values of $m_0$,  annihilation cross sections
are too small to maintain an acceptable relic density and $\Omega_\chi h^2$ is too large.
At large $m_{1/2}$,
co-annihilation processes between the LSP and the next lightest sparticle 
(in this case the $\tilde \tau$) enhance the annihilation cross section and reduce the
relic density.  This occurs when the LSP and NLSP are nearly degenerate in mass.
The dark (red) shaded region has $m_{\tilde \tau}< m_\chi$
and is excluded.   The effect of coannihilations is
to create an allowed band about 25-50 GeV wide in $m_0$ for $m_{1/2} \la
950$ GeV, or $m_\chi \la 400$ GeV, which tracks above the $m_{{\tilde \tau}_1}
 =m_\chi$ contour~\cite{efo}.  Also shown in the figure are some phenomenological
 constraints from the lack of detection of charginos \cite{LEPsusy}, or Higgses
\cite{LEPHiggs} as well as constraints from $b \to s \gamma$ \cite{bsgex}
and $g_\mu - 2$ \cite{newBNL}.  The locations of these constraints are described in the caption.

At larger $m_{1/2}, m_0$ and $\tan \beta$, the relic neutralino
density may be reduced by rapid annihilation through direct-channel $H, A$ Higgs 
bosons, as seen in Fig.~\ref{fig:UHM}(b) \cite{funnel,efgosi}.
Finally, the relic density can again be brought
down into the WMAP range at large $m_0$ 
in the `focus-point' region close the boundary where EWSB 
ceases to be possible and the lightest neutralino $\chi$
acquires a significant higgsino component \cite{fp}. The start of the focus point
region is seen in the upper left of Fig.~\ref{fig:UHM}b.

\section{Mastercode: Markov-Chain Monte-Carlo}

It is well established that Markov-Chain Monte-Carlo (MCMC) algorithms 
offer an efficient technique for sampling a large parameter space such as 
the CMSSM or its variants.  MCMC has been utilized in the Mastercode \cite{mcweb}
framework  which incorporates a code for the electroweak
observables based on~\cite{Svenetal} as well as the {\tt SoftSUSY}~\cite{Allanach:2001kg}, {\tt FeynHiggs}~\cite{FeynHiggs}, 
{\tt SuFla}~\cite{SuFla}, {\tt SuperIso}~\cite{SuperIso}, {\tt MicrOMEGAs}~\cite{MicroMegas} 
and {\tt SSARD}~\cite{SSARD} codes, using the SUSY Les Houches
Accord~\cite{SLHA}. The MCMC technique is used to sample the SUSY parameter space,
and thereby construct the $\chi^2$ probability function,
$P(\chi^2,N_{\rm dof})$. This accounts for the number of degrees of
freedom, $N_{\rm dof}$, and thus provides a quantitative measure for the
quality-of-fit such that $P(\chi^2,N_{\rm dof})$ can be used to estimate the
absolute probability with which the CMSSM describes the experimental data. 

The results of the mastercode analysis include the parameters of the
best-fit points as well as the 68 and
95\%~C.L.\ regions  
found with default implementations of the phenomenological, experimental
and cosmological constraints. These include precision electroweak data,
the anomalous magnetic moment of the muon, \gmt, 
$B$-physics observables, the bound on the lightest MSSM Higgs
  boson mass, $\Mh$, and the cold dark matter (CDM) density
inferred from astrophysical and cosmological data assuming that this is
dominated by the relic density of the lightest neutralino, $\Och$.
In addition one can include
the constraint imposed by the experimental
upper limit on the spin-independent DM scattering
cross section $\ssi$.
A purely frequentist analyses of the CMSSM
was performed in \cite{mc1,mc2,mc3,mc5,mc6,mc7}, in the NUHM1 in \cite{mc3,mc5,mc6,mc7},
and in the VCMSSM/mSUGRA in \cite{mc4,mc5,mc6}.

\begin{figure}[ht]
\hskip 1in
\begin{picture}(300,200)
  \put( 20,   20){ \resizebox{0.65\textwidth}{!}{\includegraphics{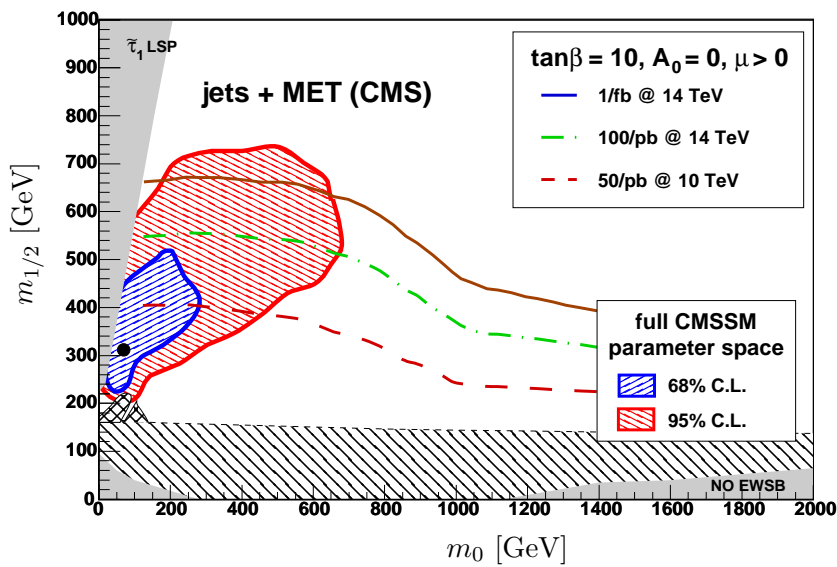}}}
  \put(170,   5){$m_0$ [GeV]}
  \put( 15,   100){\begin{rotate}{90}$m_{1/2}$ [GeV]\end{rotate}}
\end{picture}
\caption{\label{fig:MCMC}
{\it The $(m_0, m_{1/2})$ plane in the CMSSM
showing the regions favoured in a likelihood analysis
at the 68\% (blue) and 95\% (red) confidence levels~\protect\cite{mc2}. The best-fit
point is shown by the black point. Also shown are
the $5\,\sigma$ discovery contours for jet + missing $E_T$ events 
  at CMS with 1~fb$^{-1}$ at 14~TeV, 100~pb$^{-1}$ at 14~TeV and
  50~pb$^{-1}$ at 10~TeV centre-of-mass energy.}}
\end{figure}

\begin{figure*}[htb!]
{\resizebox{8cm}{!}{\includegraphics{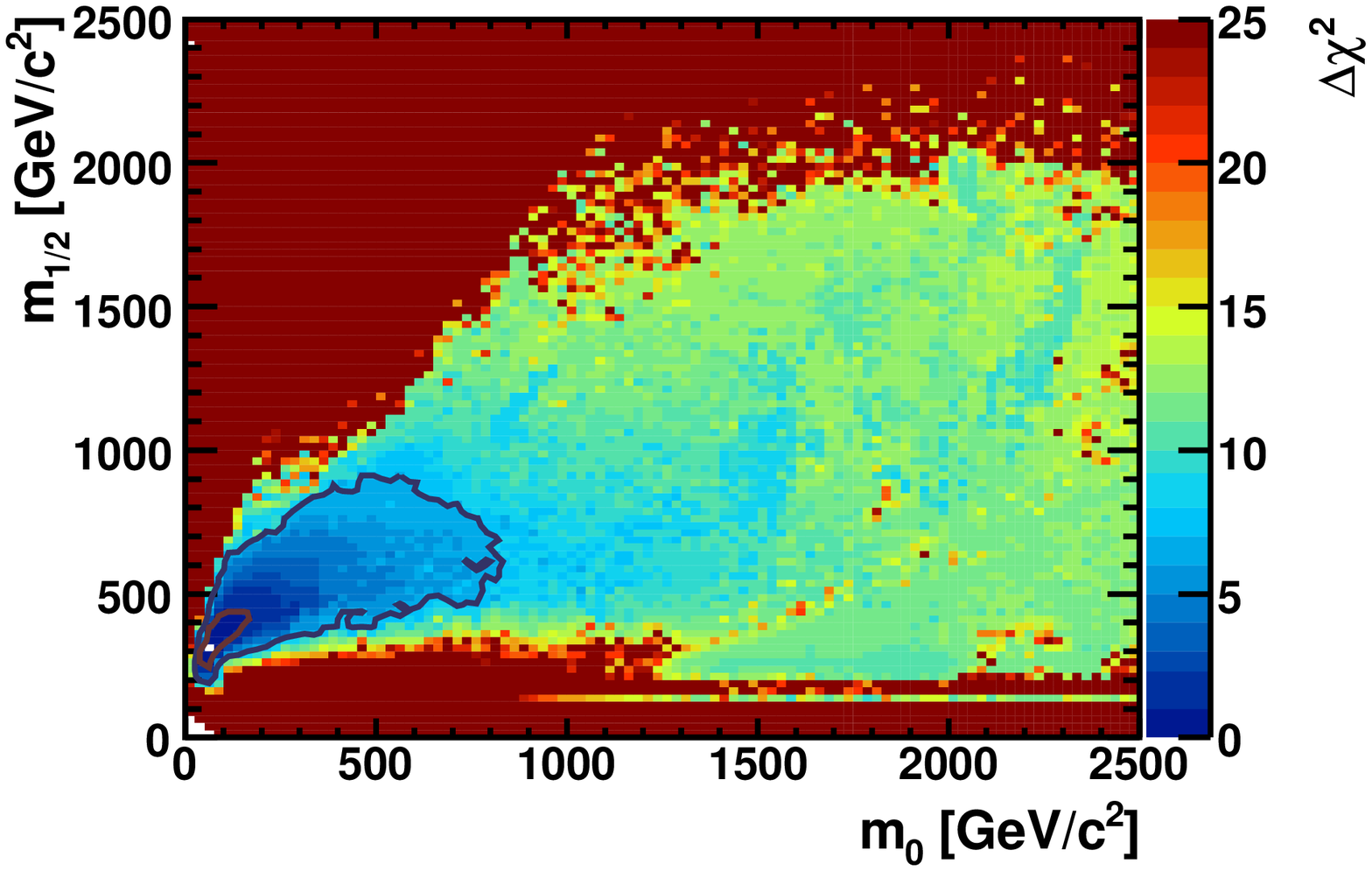}}}  
{\resizebox{8cm}{!}{\includegraphics{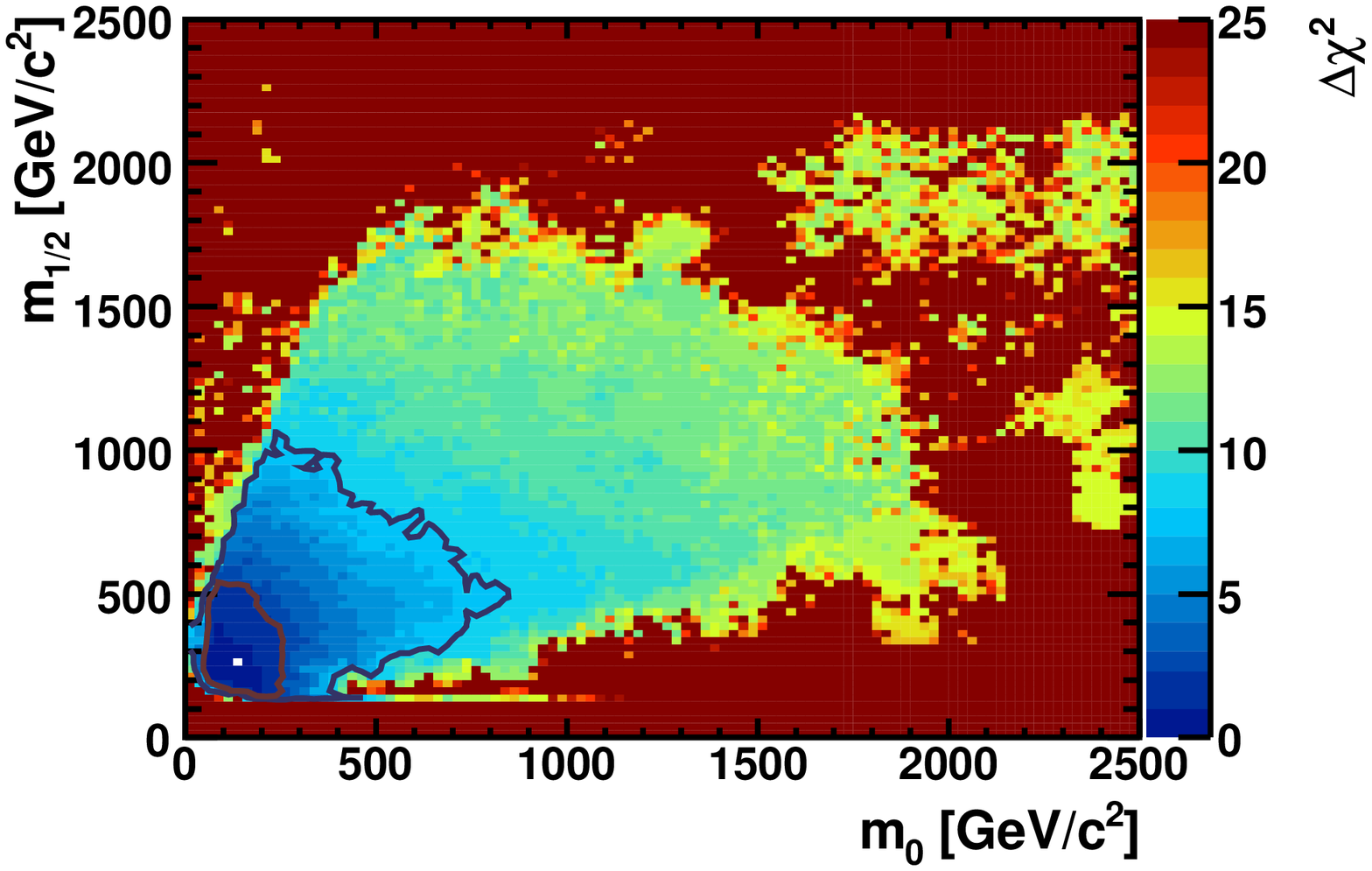}}}
\caption{\it The $\Delta\chi^2$ functions in the $(m_0, m_{1/2})$ planes for
  the CMSSM (left plot) and 
  for the NUHM1 (right plot). The best fit points are indicated by a white dot and the 68\% and 95\%
  CL contours are the light and dark curves respectively \cite{mc3}. }
\label{fig:m0m12}
\end{figure*}

In \cite{mc2}, a pre-LHC analysis of the CMSSM was performed.
The 68\% and 95\% confidence-level (C.L.) regions in the
$(m_{1/2}, m_0)$ plane of the CMSSM are shown in
Fig.~\ref{fig:MCMC}. Also shown for comparison are the physics
reaches of ATLAS and CMS with 1/fb of integrated luminosity~\cite{:1999fr,Ball:2007zza}. 
 The likelihood
analysis assumed $\mu > 0$, as motivated by the sign of the
apparent discrepancy in $g_\mu - 2$, but sampled all values of $\tan \beta$
and $A_0$: the experimental sensitivities were estimated
assuming $\tan \beta = 10$ and $A_0 = 0$, but are probably not
very sensitive to these assumptions. The global maximum of the
likelihood function (indicated by the black dot) is at
$m_{1/2} = 310$~GeV,
$m_0 = 60$~GeV, $A_0 = 240$~GeV, $\tan \beta = 11$ and
$\chi^2/N_{dof} = 20.4/19$ (37\% probability). Note that the best-fit point lies
well within the LHC discovery range, as does the 68\%  C.L. region. 
As we wil see, by the end of 2011, the LHC has met this reach (at 7 TeV center of mass energy)
and as sparticles have yet to be discovered, this region is mostly excluded at 95\% CL
(see sections 4 and 5 below).
A more detailed view of the $\Delta \chi^2$ function for the CMSSM is shown in
Fig.~\ref{fig:m0m12}. For other pre-LHC results see \cite{pre-LHC}.

\section{Results for the NUHM1} 

\begin{figure}[htb]
{\resizebox{8cm}{!}{\includegraphics{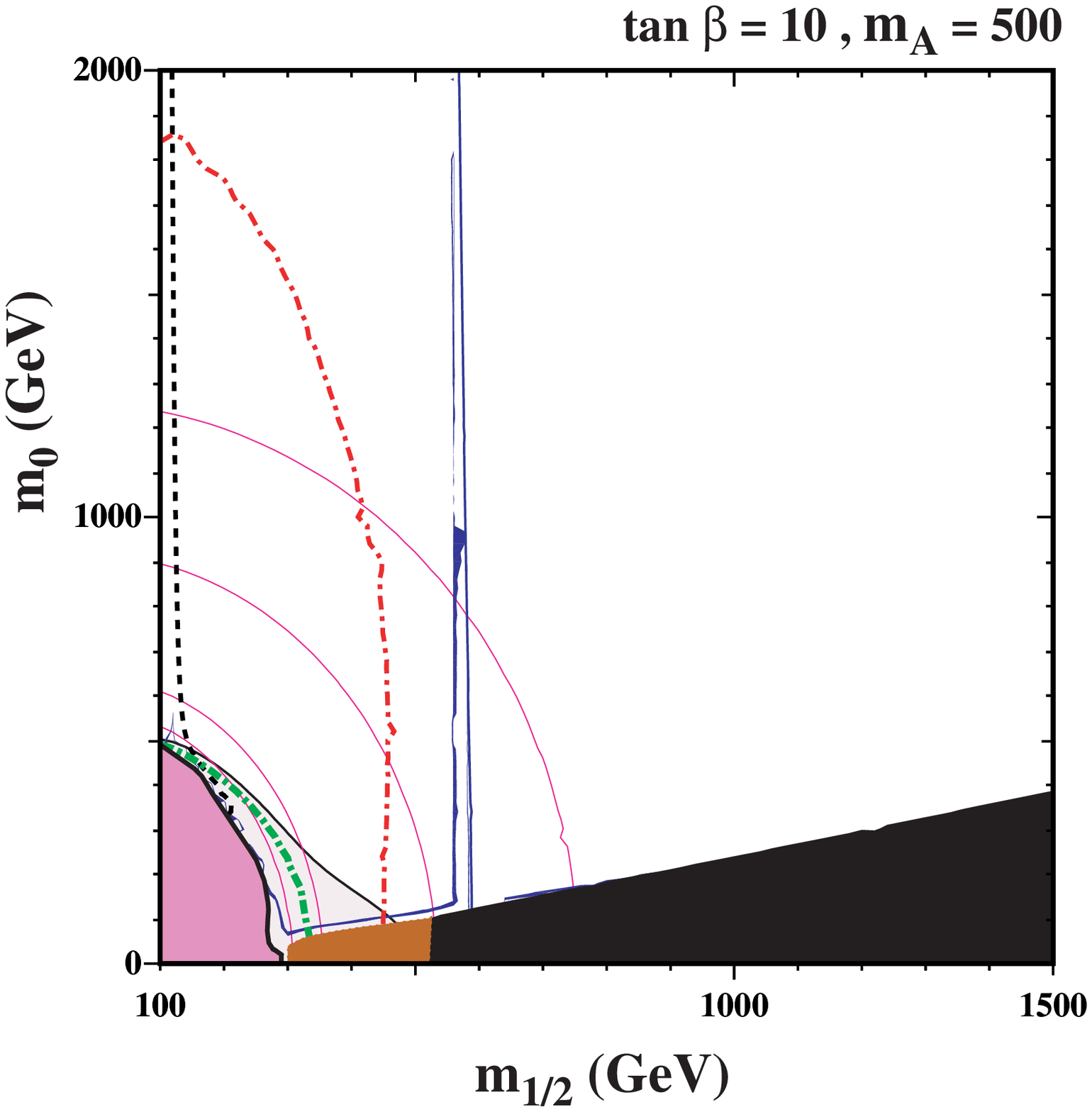}}}
 {\resizebox{8cm}{!}{\includegraphics{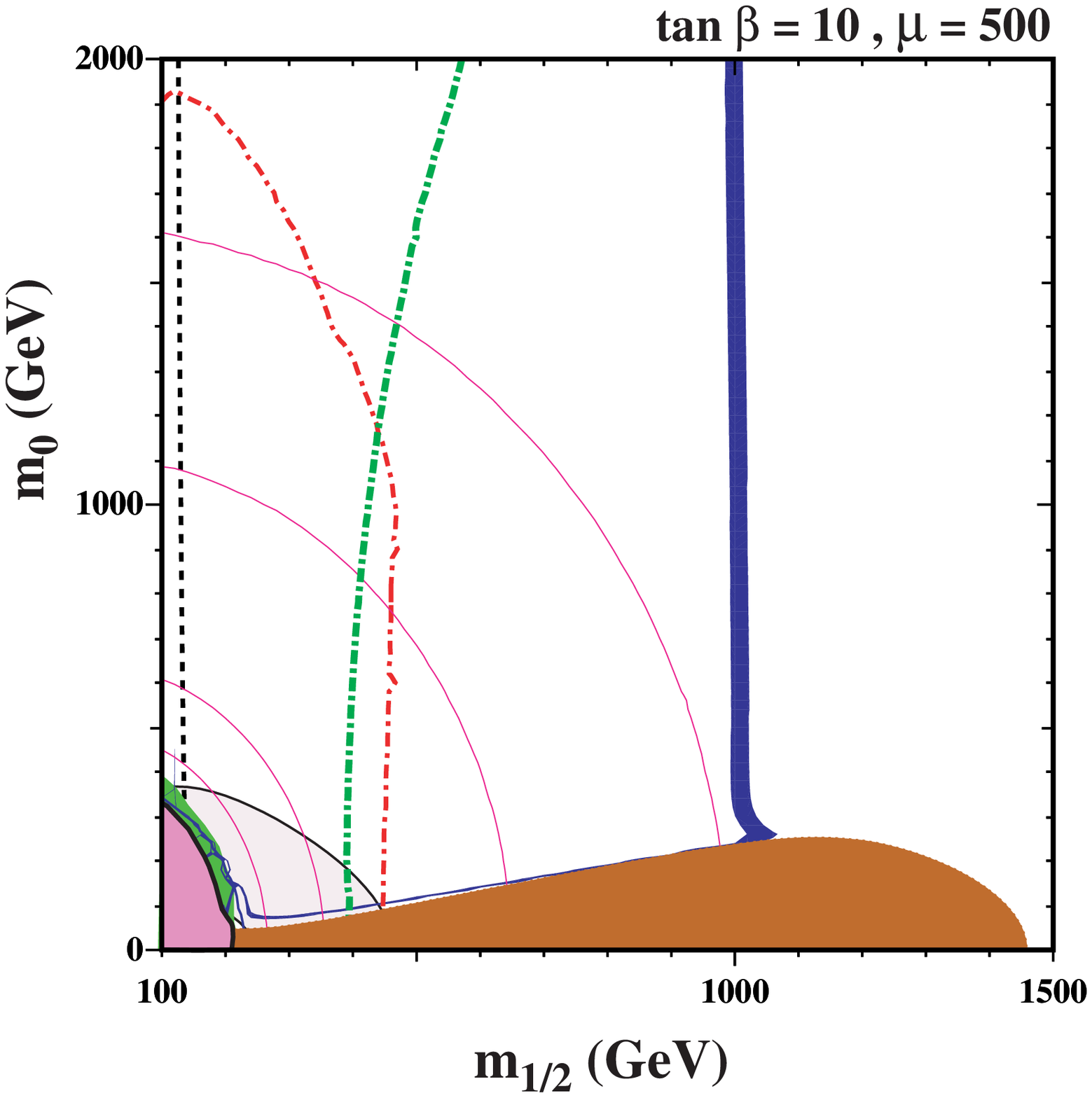}}}
\caption{\it The $(m_{1/2}, m_0)$ plane in the NUHM for $\tan \beta = 10$,  a) with fixed 
$m_A = 500$~GeV  and b) with fixed $\mu = 500$~GeV~\protect\cite{nuhm1-2}. }
\label{fig:NUHM}       
\end{figure}

The cosmologically preferred regions move around
in the $(m_{1/2},m_0)$ plane if one abandons the universality assumptions of the
CMSSM. As discussed above, if one allows the supersymmetry-breaking contributions
to the Higgs masses to be non-universal (NUHM), the rapid-annihilation WMAP 
`strip' can appear at different values of $\tan \beta$ and $m_{1/2}$, as seen in
Fig.~\ref{fig:NUHM}~\cite{nuhm1-2}. In the left panel, we show an
NUHM1 $(m_{1/2},m_0)$ plane
for $\tan \beta = 10, A_0 = 0$ and $\mu > 0$
with $m_A=500$ GeV, and $\mu$ calculated
using (\ref{eq:mu}). In addition to the constraints discussed above,
we also plot contours of $\mu = 300$, 500, 1000, and 1500 GeV (light pink). 
The  thick green dot-dashed contour
tracks the CMSSM parameters in the NUHM1 $(m_{1/2},m_0)$ plane.
The
most prominent departure from the CMSSM is that the EWSB requirement 
constrains the plane at low 
$m_0$ rather than at large $m_0$. In this region (below the CMSSM
contour), the fixed value $m_A$ is larger than its corresponding value in the CMSSM, resulting
in correspondingly larger $m_1^2$ and $m_2^2$ (smaller $|m_2^2|$, since $m_2^2 < 0$). We see from
(\ref{eq:mu}) that,  the effect is to drive $\mu^2$ smaller, and eventually negative. 
The stau LSP exclusion regions are qualitatively similar to those in the CMSSM,
however there is a (black shaded) region
of the plane where the lighter selectron is the LSP.
In this case, the co-annihilation region connects the analogue of the focus point region with small
$\mu$ discussed above and heavy Higgs funnel region which now exists at $\tan \beta = 10$ at
$m_{1/2} \approx 550$ GeV and extends to large $m_0$. Some of the co-annihilation region also
extends to larger $m_{1/2}$. 

%

In the right panel of  Fig.~\ref{fig:NUHM}, we show the NUHM1 $(m_{1/2},m_0)$ plane with
$\mu=500$~GeV  and $m_A$ calculated
using (\ref{eq:muB}).
At first glance, the $(m_{1/2}, m_0)$ plane with fixed $\mu$ has
some similarities with those with fixed $m_A$. There are excluded regions at
very low $(m_{1/2}, m_0)$ where the pseudoscalar Higgs mass squared is
negative, corresponding to the absence of electroweak symmetry breaking,
surrounded by four contours of fixed $m_A = 300$, 500, 1000, and 1500~GeV. 
At small values of $m_0$, extending out to large $m_{1/2}$, there are
excluded $\tilde \tau$-LSP regions resembling those in the CMSSM.
$b \to s \gamma$ excludes strips near the EWSB boundary.
In addition to the 
coannihilation strip close to the $\tilde \tau$-LSP boundary, another strip
close to the EWSB boundary, and the
curved rapid-annihilation funnels that appear at low $m_A$, with
strips of good relic density forming the funnel walls, there is 
a fourth, near-vertical strip,
where the relic density is brought down into the WMAP range because of the large
mixing between the bino and Higgsino components in the LSP.
 For smaller $m_{1/2} < 500$~GeV,
the LSP is almost pure bino, and the relic density is too large except in the narrow
strips mentioned previously. On the other hand, for larger $m_{1/2} > 1000$~GeV, the LSP is
almost pure Higgsino, and the relic density falls below the WMAP 
range. It is also this change in the nature of the LSP that causes the
boundary of the $\tilde \tau$LSP region to drop. Since the
$\tilde \tau$ mass is affected only minimally by the value of $\mu$, we find
that $\tilde \tau$-LSP region terminates at some value of $m_{1/2}$ related
primarily to $\mu$. At large $m_0$ in panel (b) 
of Fig.~\ref{fig:NUHM}, 
it is only in the `crossover' strip  that the relic density falls within the WMAP range.
The CMSSM contour 
is a roughly vertical thick green dot-dashed line, the position of which
is determined by the value of $m_A$ that one would find from the
electroweak vacuum conditions in the
standard CMSSM.

The results of the Mastercode (pre-LHC) analysis for the NUHM is shown in the
right panel of Fig.~\ref{fig:m0m12}.
The corresponding parameters of
the best-fit NUHM1 point are $m_0 = 150 \gev$, $m_{1/2} = 270 \gev$,
$A_0 = -1300 \gev$, $\tb = 11$ and
$m_{h_1}^2  = m_{h_2}^2 = - 1.2 \times 10^6 \gev^2$ or, equivalently,
$\mu = 1140 \gev$, yielding
$\chi^2 = 18.4$ (corresponding to a similar fit probability to the CMSSM)
and $\Mh = 120.7 \gev$. 

\begin{figure*}[htb!]
\resizebox{8.0cm}{!}{\includegraphics{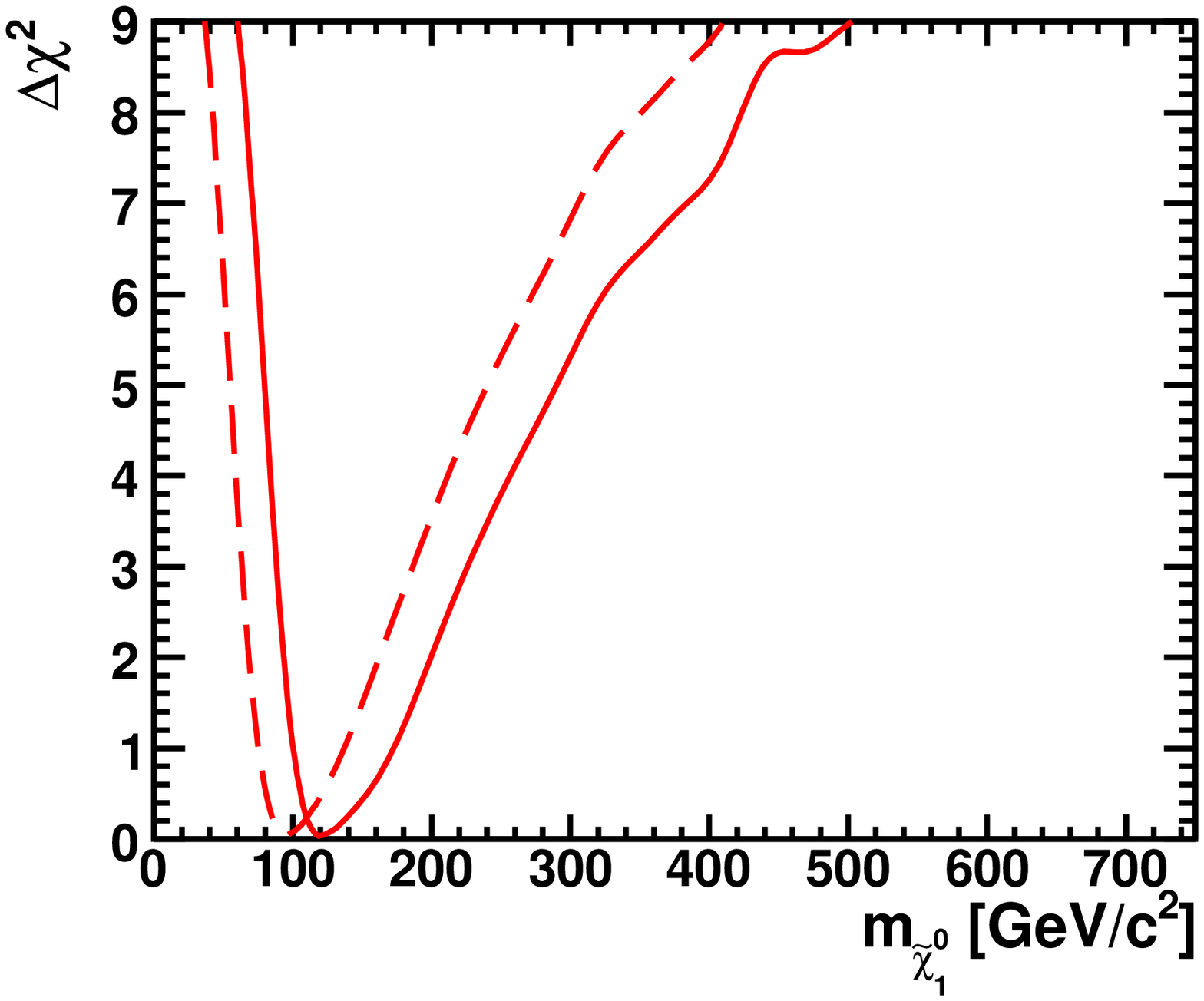}}
\resizebox{8.0cm}{!}{\includegraphics{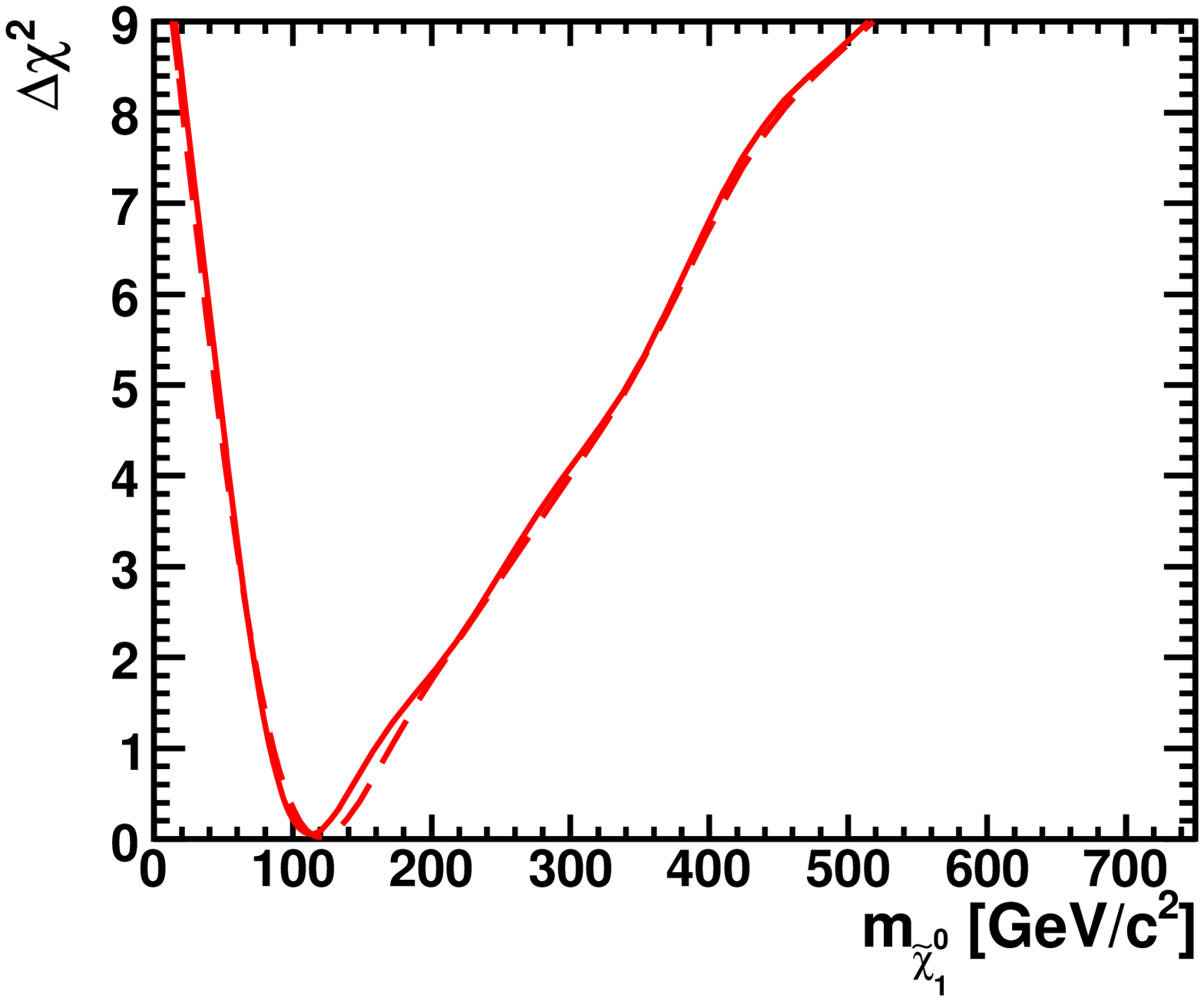}}
\vspace{-1em}
\caption{\it The likelihood functions for $\mneu{1}$ in the CMSSM (left)
and in the NUHM1 (right), both with (solid lines) and without
(dashed lines) the LEP constraint on $\Mh$ \cite{mc3}.
}
\label{fig:mchi}
\end{figure*}

It is also possible to extract 1D likelihood functions for essentially any parameter of interest.
Here, we show in Fig.~\ref{fig:mchi} 
the $\Delta \chi^2$ function for the lightest neutralino mass in
both the CMSSM and NUHM1.
The left panel of Fig.~\ref{fig:mchi} displays the likelihood function
in the CMSSM. 
The solid line shows the result obtained when
incorporating the LEP Higgs limit, while the dashed line corresponds to
the case where the LEP Higgs constraint is removed.
There is a sharp rise in the likelihood function at low values of
$\mneu{1}$, which is caused by the limits from the direct searches for
SUSY particles, but receives also contributions from
\bsg\ and other constraints. This sharp rise in the likelihood function 
persists when the LEP Higgs constraint is removed,
but is shifted towards slightly lower values of $\mneu{1}$ in that
case.
The right panel of Fig.~\ref{fig:mchi}
shows the likelihood function for $\mneu{1}$ in the NUHM1, again with
and without the LEP $\Mh$ constraint imposed.

\section{Early Results from the LHC}

As 2010 came to a close, we obtained the first results from the LHC on 
sparticle searches. These include
several new constraints on SUSY using an integrated luminosity
of $\sim 35$/pb of data at 7~TeV. ATLAS has published the results of
a search in multijet + $\ETslash$ channels (ATLAS 0L)~\cite{ATLAS0l}
that has greater sensitivity in some regions to the types of gluino and squark 
pair-production events expected in the supersymmetric models discussed
here than did the earlier ATLAS 1L search~\cite{ATLAS1l}, and has also released results
obtained by combining the one- and zero-lepton searches~\cite{ATLAScombined}.
CMS has announced results from two other searches
in multijet + $\ETslash$ channels that improve the CMS $\alpha_T$
sensitivity also to gluino and squark 
production in the models discussed here \cite{cms0l-aT,MHT}.

In addition, the XENON100 Collaboration has recently released results from a
search for direct spin-independent dark matter scattering with 100.9
live days of data using a fiducial 
target with a mass of 48~kg~\cite{Xenon100new}. As we see later, this provides 
constraints on the parameter spaces of supersymmetric models that complement
those provided by collider experiments.

\begin{figure*}[htb!]
\resizebox{8cm}{!}{\includegraphics{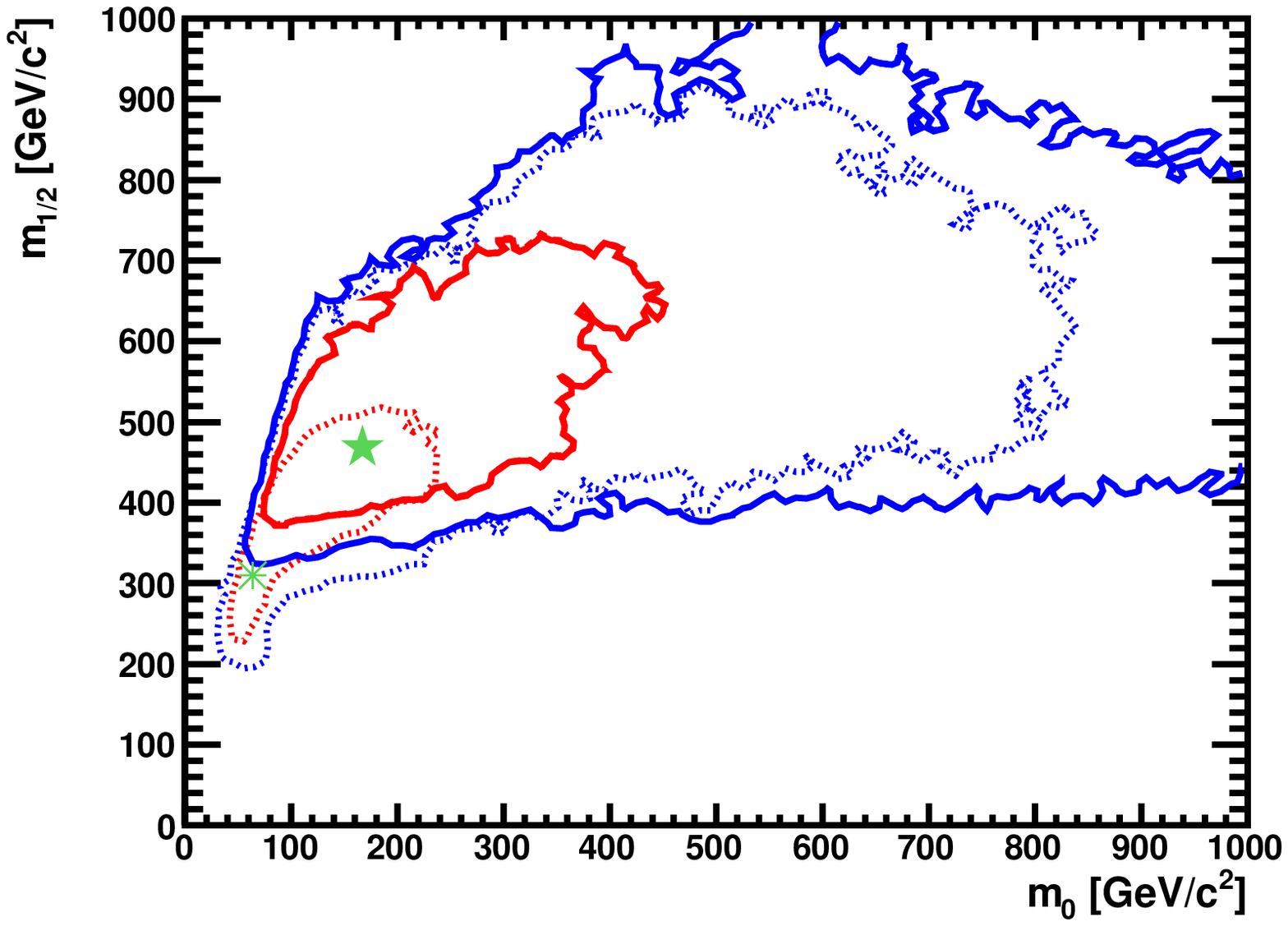}}
\resizebox{8cm}{!}{\includegraphics{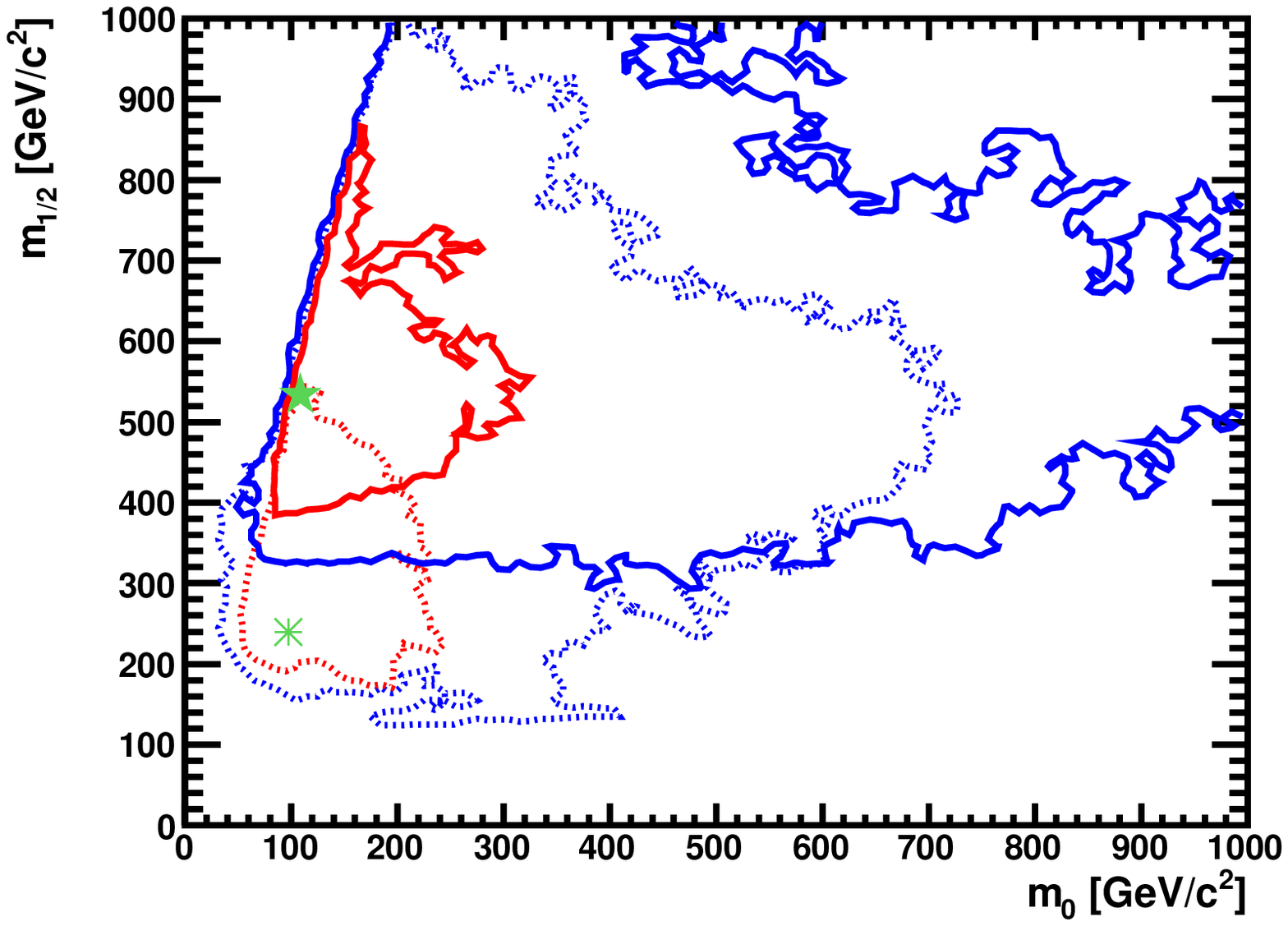}}
\vspace{-1cm}
\caption{\it The $(m_0, m_{1/2})$ planes in the CMSSM (left), and the
  NUHM1 (right). In each plane, the best-fit point after incorporation of the
  2010 LHC and XENON100 constraints is indicated by a filled green star, and the
  pre-LHC fit by an open star. The 68 and 95\% CL regions are indicated
  by red and blue contours, respectively, the solid lines including the
  2010 LHC and XENON100 data, 
  and the dotted lines showing the pre-LHC fits \cite{mc5,mc6}. 
}
\label{fig:lhcm0m12}
\end{figure*}

The impact of the 35/pb LHC data on the CMSSM parameter space is rather dramatic \cite{mc5,mc6,mc7}
(see also \cite{post-LHC}).
In Fig. \ref{fig:lhcm0m12}, we display the planes for the
CMSSM (left) and NUHM1 (right) driven by the ATLAS 0L and CMS MHT constraints but also
taking into account the other 2010 LHC constraints, as well as the XENON100 constraint \cite{mc6}. 
The best-fit points are shown as green stars and 68 and 95\% CL regions are shown as
red and blue lines,
respectively and correspond to $\Delta \chi^2 = 2.3$ (red) and 5.99 (blue)
relative to the best fit points. The pre-LHC results, taken from \cite{mc5}, are displayed
as `snowflakes' and dotted lines, the post-2010-LHC/XENON100 results are
displayed as full stars and solid lines \cite{mc6}.

In the CMSSM and 
the NUHM,  the direct 2010 LHC constraints push the
best-fit values of $m_{1/2}$ to significantly higher values, as well as
their 68 and 95\% CL ranges.  The effect of the LHC on the best-fit values
of $m_0$ is smaller, though there is a significant increase in the
CMSSM that is correlated with the increase in $m_{1/2}$. 
The positions of the best fit points pre/post LHC are collected in Table 1.
The total value of $\chi^2$ and the number of degrees of freedom (dof)
along with the fit probably (p-value) are also given.  The final column
shows the predicted value of the Higgs mass at the best fit point.
Note there is a slight shift in the best fit points for the pre-LHC data
due to changes in the input data used in the Mastercode analysis.
LHC2010 includes the 35/pb LHC data as well as the XENON100 data.
While one would expect two additional dof's, one
dof  (the Higgs mass constraint) is lost since the best fit point is pushed
past the previous LEP sensitivity.

\begin{table*}[!tbh!]
\renewcommand{\arraystretch}{1.5}
\begin{center}
\begin{tabular}{|c||c|c|c|c|c|c||c|c|} \hline
Model & Minimum & Prob- & $m_{1/2}$ & $m_0$ & $A_0$ & $\tb$ & $\Mh$ (GeV) \\
      & $\chi^2$/d.o.f.& ability & (GeV) & (GeV) & (GeV) & & (no LEP)\\ 
\hline \hline
CMSSM  pre-LHC 
    & 21.5/20 & {37\%} & $360$ & $90$ 
    & $-400$ & $15$ & {$111.5$}\\
LHC 2010  &  25.2/21  &   24\%    &  $470$  & $170$  
      &  $-780$  &  $22$  &   115.7 \\
LHC$_{\rm 1/fb}$     
    & 28.8/22 & {15\%} & $780$ & $450$ 
    & $-1100$ & $41$ & {$119.1$}\\
\gmt\ neglected &  
    21.3/{20}  &   {38}\%    &  $2000$  & $1050$  
    & $430$ &  $22$  &   {$124.8$} \\
\hline
NUHM1 pre-LHC    
    & 20.8/18 & 29\% & $340$ & $110$ 
    & $520$ & $13$ & {$118.9$} \\
    LHC 2010  &  24.5/20  &  22\%    &  $530$  &  $110$  
      &  $ -370$  & $27$  &  117.9 \\
LHC$_{\rm 1/fb}$     
    & 27.3/21 & 16\% & $730$ & $150$ 
    & $-910$ & $41$ & {$118.8$} \\
\gmt\  neglected 
    &  20.3/{19}  &  {38}\%    &  $2020$  &  $1410$  
    &  $ 2580$  & $48$  &  {$126.6$} \\
\hline
\end{tabular}
\caption{\it Comparison of the best-fit points found in the CMSSM and NUHM1 pre-LHC (including the 
upper limit on \bmm\ available then), the LHC 2010 result (including XENON100) 
and with the LHC$_{\rm 1/fb}$ data set (also including the XENON100 constraint)
using the standard implementations of the \gmt\ constraint, followed by the case dropping \gmt. The
predictions for $\Mh$ do not include the constraint from the direct LEP Higgs search.
}
\label{tab:bestfits}
\end{center}
\end{table*}

\section{The LHC @ 1/fb}

\begin{figure*}[htb!]
\resizebox{8.7cm}{!}{\includegraphics{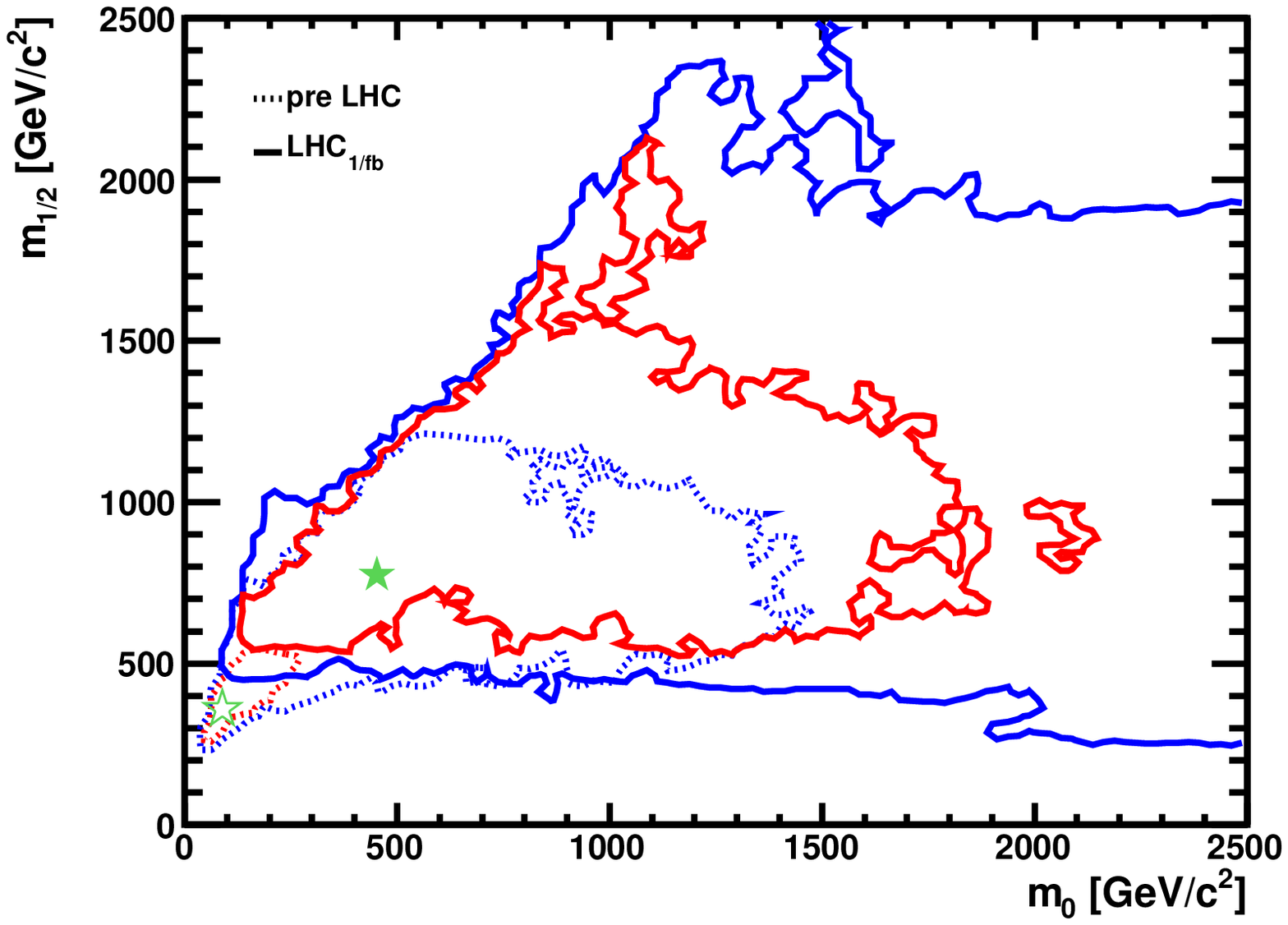}}
\resizebox{8.7cm}{!}{\includegraphics{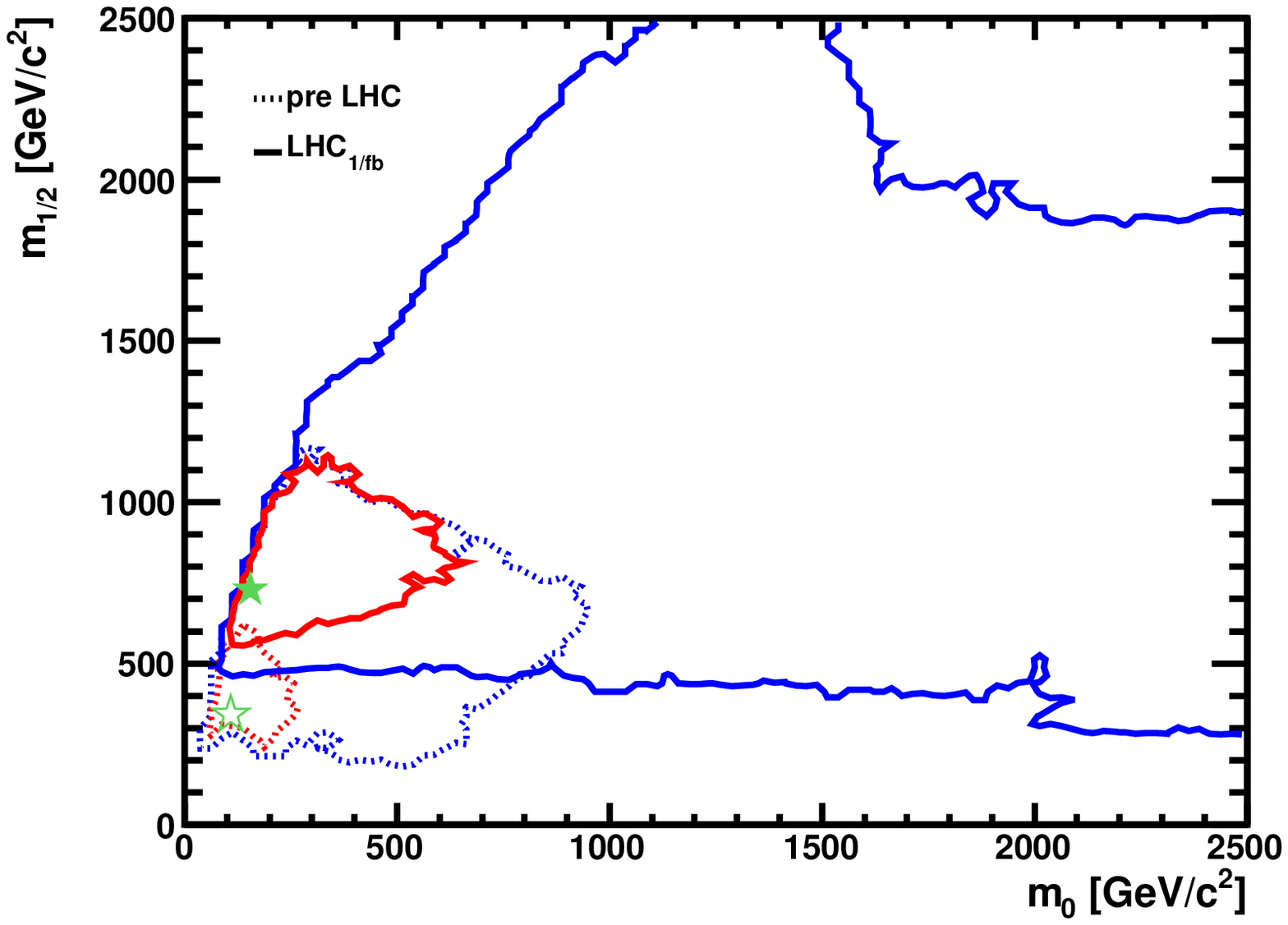}}
\vspace{-1cm}
\caption{\it The $(m_0, m_{1/2})$ planes in the CMSSM (left) and the
  NUHM1 (right). In each plane, the best-fit point after incorporation of the
  LHC$_{\rm 1/fb}$ constraints is indicated by a filled green star, and the
  pre-LHC fit~\protect\cite{mc4} by an open star. The 68 and 95\% CL regions are indicated
  in red and blue, respectively, the solid lines including the
  LHC$_{\rm 1/fb}$ data and the dotted lines showing the pre-LHC fits \cite{mc7}.
}
\label{fig:6895}
\end{figure*}

By mid 2011, the LHC results for sparticle searches based on 1/fb of analyzed data
were released by the ATLAS~\cite{ATLASsusy,ATLASHA}, CMS~\cite{CMSsusy,CMSHA,CMSbmm} 
and LHCb Collaborations~\cite{LHCbbmm}.
The absences of signals
in the jets + $\ETslash$ searches disfavour the ranges of the model mass 
parameters $(m_0, m_{1/2})$ that had been favoured in our previous analyses 
of the CMSSM and NUHM1~\cite{mc5,mc6}, and our current best fits have $m_0 \sim 150$ to 450~GeV
and $m_{1/2} \sim 750 \gev$. Reconciling these larger values of $(m_0, m_{1/2})$
with \gmt\ favours values of $\tan \beta \sim$ 40, though with a large uncertainty.
Fig. \ref{fig:6895} shows the positions of the 68 and 95\% CL contours with
solid curves corresponding to the 1/fb data as compared with the pre-LHC (and pre-XENON100) results (dashed) \cite{mc7}. 

The positions of the best fit points at 1/fb are tabulated in Table \ref{tab:bestfits}.
We now find that the $p$-value for the CMSSM best-fit point is now
$\sim 15$\%, and that for the NUHM1 
is $\sim 16$\%. On the other hand, if the \gmt\
constraint is dropped much larger regions of the
$(m_0, m_{1/2})$ and other parameter planes are allowed at the 68 and 95\% CL,
and these $p$-values increase to 38\% in both models. For comparison, the p-value for the 
Standard Model (including \gmt) is 9\%.

\begin{figure*}[htb!]
\resizebox{9cm}{!}{\includegraphics{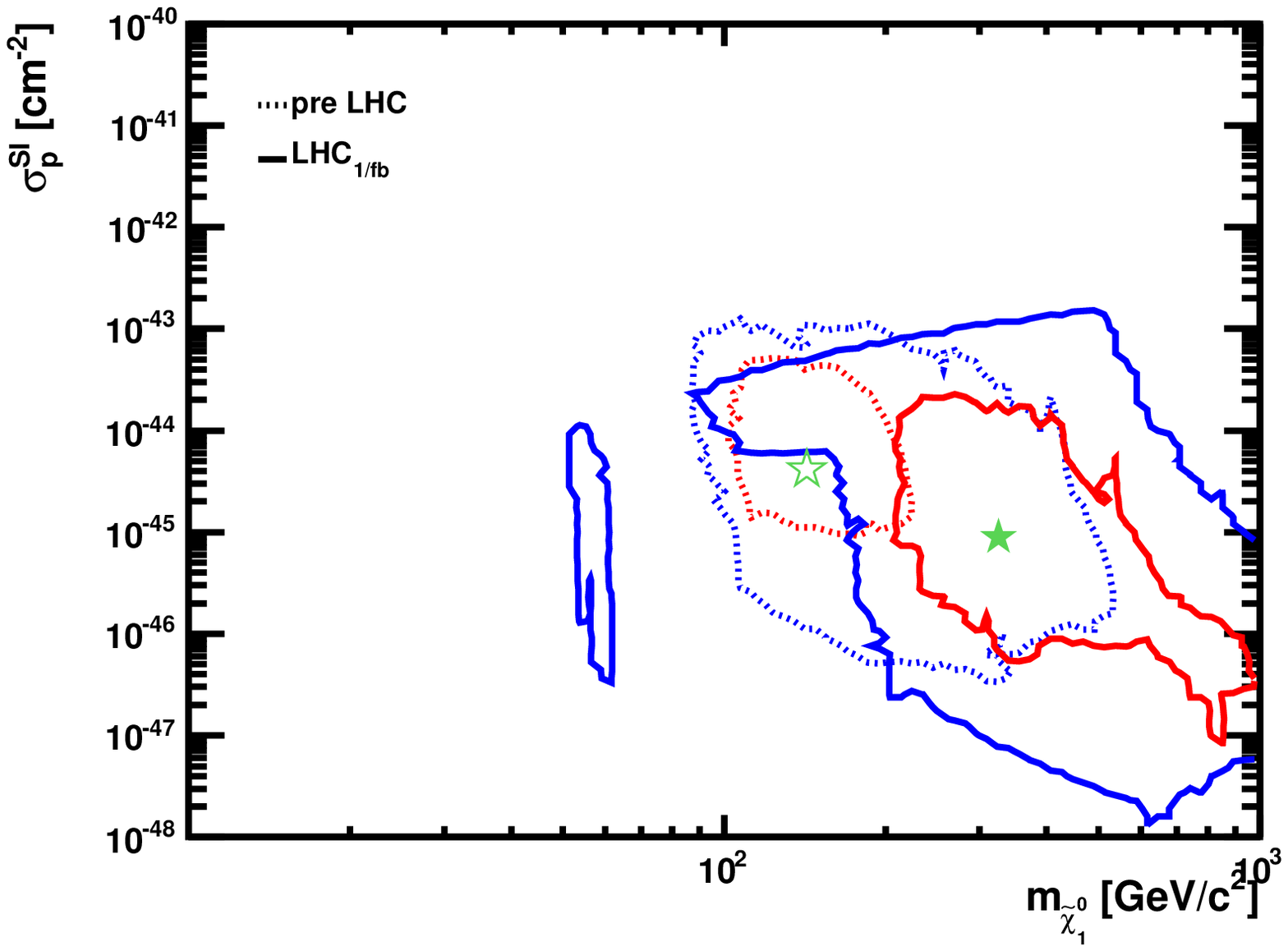}}
\resizebox{9cm}{!}{\includegraphics{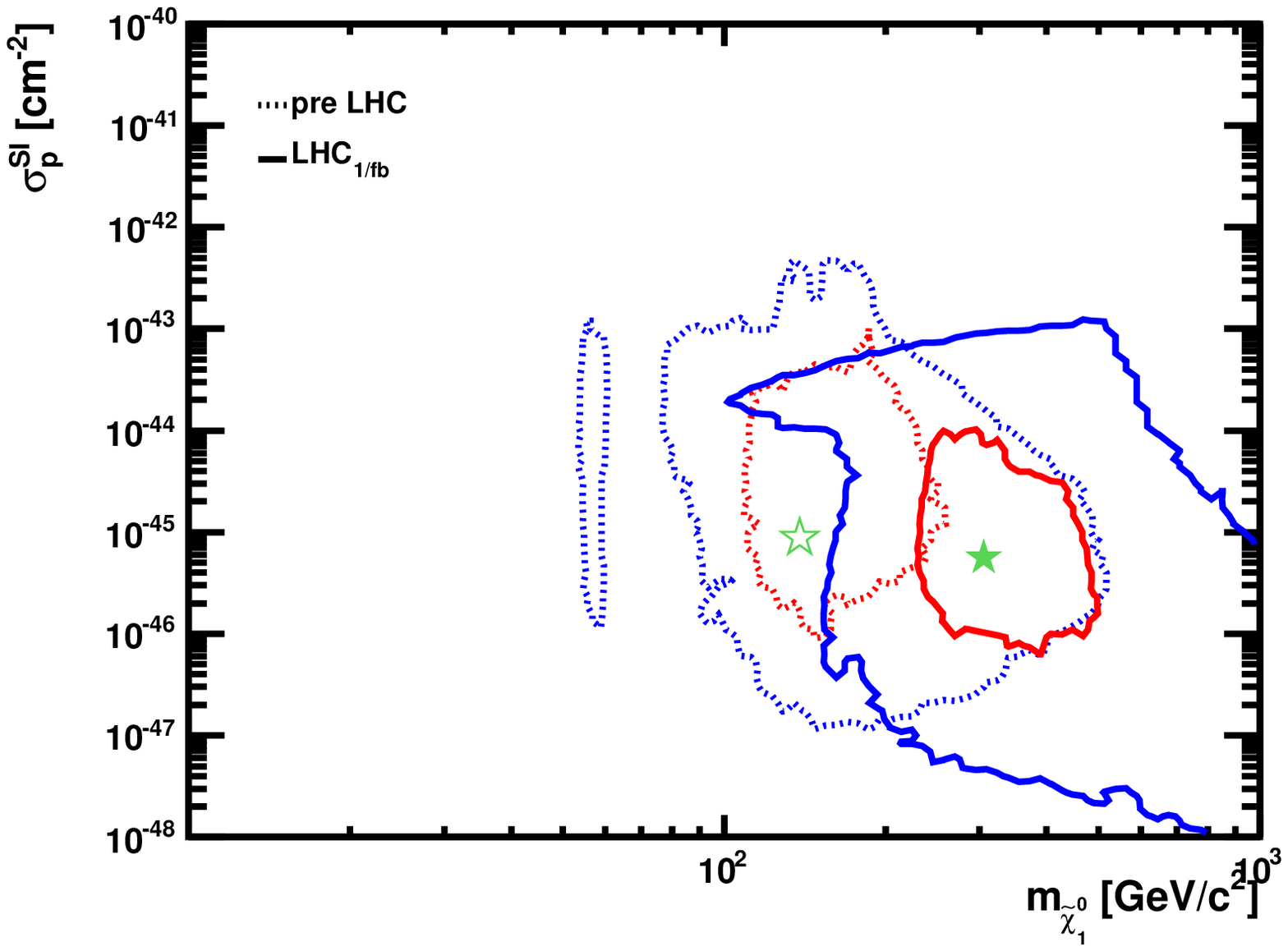}}
\vspace{-1cm}
\caption{\it The 68\% and 95\%~CL contours (red and blue, respectively)
in the CMSSM (left) and the NUHM1 (right). The solid lines are for
fits including the XENON100~\protect\cite{Xenon100new} and LHC$_{\rm 1/fb}$ data, whereas the dotted lines 
include only the pre-LHC data~\protect\cite{mc7}.
}
\label{fig:ssi}
\end{figure*}

In Fig.~\ref{fig:ssi} we show the 68\% and 95\%~CL contours in the $(\mneu{1}, \ssi)$ planes
for the CMSSM (left) and the NUHM1 (right). The solid lines are based on
our global fits including the LHC$_{\rm 1/fb}$ constraints, whereas the dotted lines
correspond to our previous fits using the pre-LHC constraints. In both
cases, we assume $\Sigma_{\pi N} = 50 \pm 14$ MeV~\cite{sigma}~\footnote{We recall
the sensitivity of predictions for $\ssi$\ to the uncertainty in $\Sigma_{\pi N}$~\cite{mc6}.}, 
and we include with the LHC$_{\rm 1/fb}$ data the
XENON100 constraint on $\ssi$~\cite{Xenon100new}. We see that the LHC$_{\rm 1/fb}$ data
tend to push $\mneu{1}$ to larger values, and that these are correlated with lower values
of $\ssi$, though with best-fit values still $\sim 10^{-45}$~cm$^2$ \cite{mc7}. 

\section{The Higgs Search}

At the close of 2011,  the ATLAS and CMS Collaborations have released their
official combination of the searches for a SM Higgs boson with the first
$\sim 1 - 2.3$/fb of LHC luminosity at $E_{\rm cm} =
7$~TeV~\cite{ATLAS+CMS}. The combination excludes a SM
Higgs boson with a mass between 141 and 476~GeV. Additionally, the ATLAS and
CMS Collaborations have presented preliminary updates of their results with
$\sim 5$/fb of data~\cite{Dec13}. These results may be compatible with a SM-like Higgs
boson around $\Mh \simeq 125 \gev$, though CMS also report an
excess at $\Mh \simeq 119 \gev$ in the $ZZ^*$ channel.

It is interesting to note that based on the LHC 1/fb results, we have the predictions of the
Higgs mass as seen in Table \ref{tab:bestfits} of 119 and ~125 GeV depending on whether
\gmt\ is included in the analysis \cite{mc7}.  Specifically, the 1/fb fits  included $\Mh = 119.1^{+3.4}_{-2.9} \gev$ in the CMSSM
and $\Mh = 118.8^{+2.7}_{-1.1} \gev$ in the NUHM1 (which should be combined
with an estimated theory error $\Delta \Mh = \pm 1.5 \gev$). 
These two fits are based solely on the Higgs-{\it independent} searches
including the 
\gmt\ constraint, i.e., they do not 
rely on the existing limits from LEP~\cite{LEPHiggs,Schael:2006cr},
the Tevatron~\cite{TevHiggs}, or the LHC~\cite{ATLASHA,CMSHA}.
These predictions increase to $\Mh = 124.8^{+3.4}_{-10.5} \gev$ in the CMSSM
and $126.6^{+0.7}_{-1.9} \gev$ in the NUHM1 if the \gmt\ constraint is dropped.

\begin{figure*}[htb!]
\resizebox{8.6cm}{!}{\includegraphics{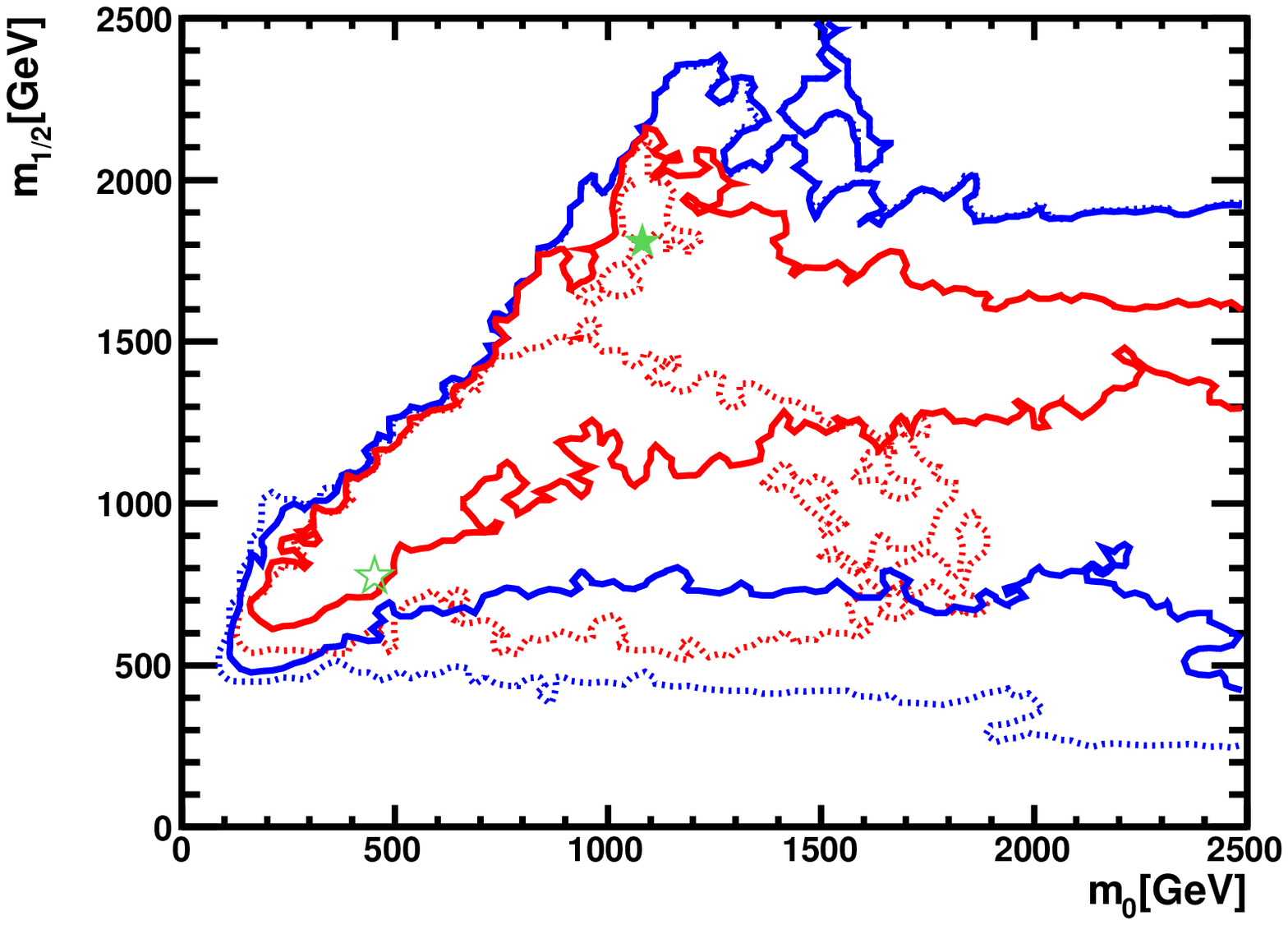}}
\resizebox{8.6cm}{!}{\includegraphics{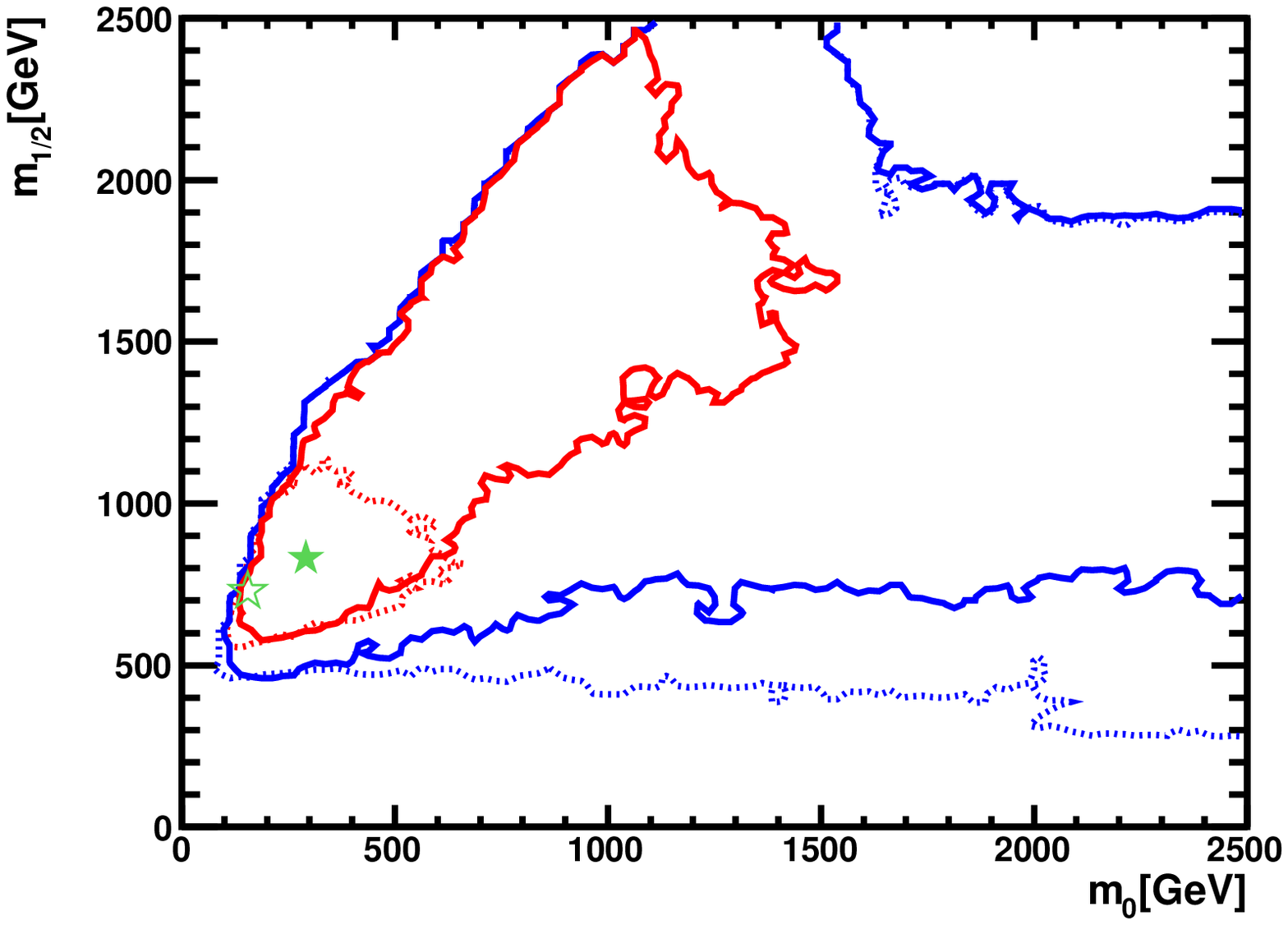}}
\caption{\it The $(m_0, m_{1/2})$ planes in the CMSSM (left) and the
  NUHM1 (right). The  68 and 95\% CL regions
  are indicated in red and blue, respectively, the solid lines including
  the hypothetical LHC measurement $\Mh = 125 \pm 1 \gev$
  and allowing for a theoretical error $\pm 1.5 \gev$ \cite{mc7.5},
  and the dotted lines showing the contours
  found previously in~\protect\cite{mc7} without this $\Mh$ constraint.
  Here the open green stars denote the pre-Higgs best-fit
  points, whereas the solid green stars indicate the new
  best-fit points. 
}
\label{fig:125}
\end{figure*}

If indeed, the LHC has seen the Higgs at 125$\pm$1 (119$\pm$1) GeV, there are
rather dramatic consequences for the supersymmetric parameter space \cite{mc7.5} (see also \cite{post-mh}).
Since in the CMSSM and NUHM1 the radiative corrections 
contributing to the value of $\Mh$ are sensitive primarily to $m_{1/2}$
and $\tb$, and only to a lesser extent to $m_0$, we expect that the primary
effect of imposing the $\Mh$ constraint should be to affect the
preferred ranges of $m_{1/2}$ and $\tb$, with a lesser effect on the preferred 
range of $m_0$. 
This effect is indeed seen in both panels of Fig.~\ref{fig:125}. 
We see that the 68\% CL ranges of $m_{1/2}$ extend to
somewhat larger
values and with a wider spread than the pre-Higgs results, particularly in the NUHM1.
However, the NUHM1 best-fit value of $m_{1/2}$ remains at a relatively low value
of $\sim 800 \gev$, whereas the best-fit value of $m_{1/2}$ in the CMSSM moves to
$\sim 1900 \gev$. This jump reflects the flatness of the likelihood function
for $m_{1/2}$ between $\sim 700 \gev$ and $\sim 2 \tev$.

\begin{figure*}[htb!]
\resizebox{8.6cm}{!}{\includegraphics{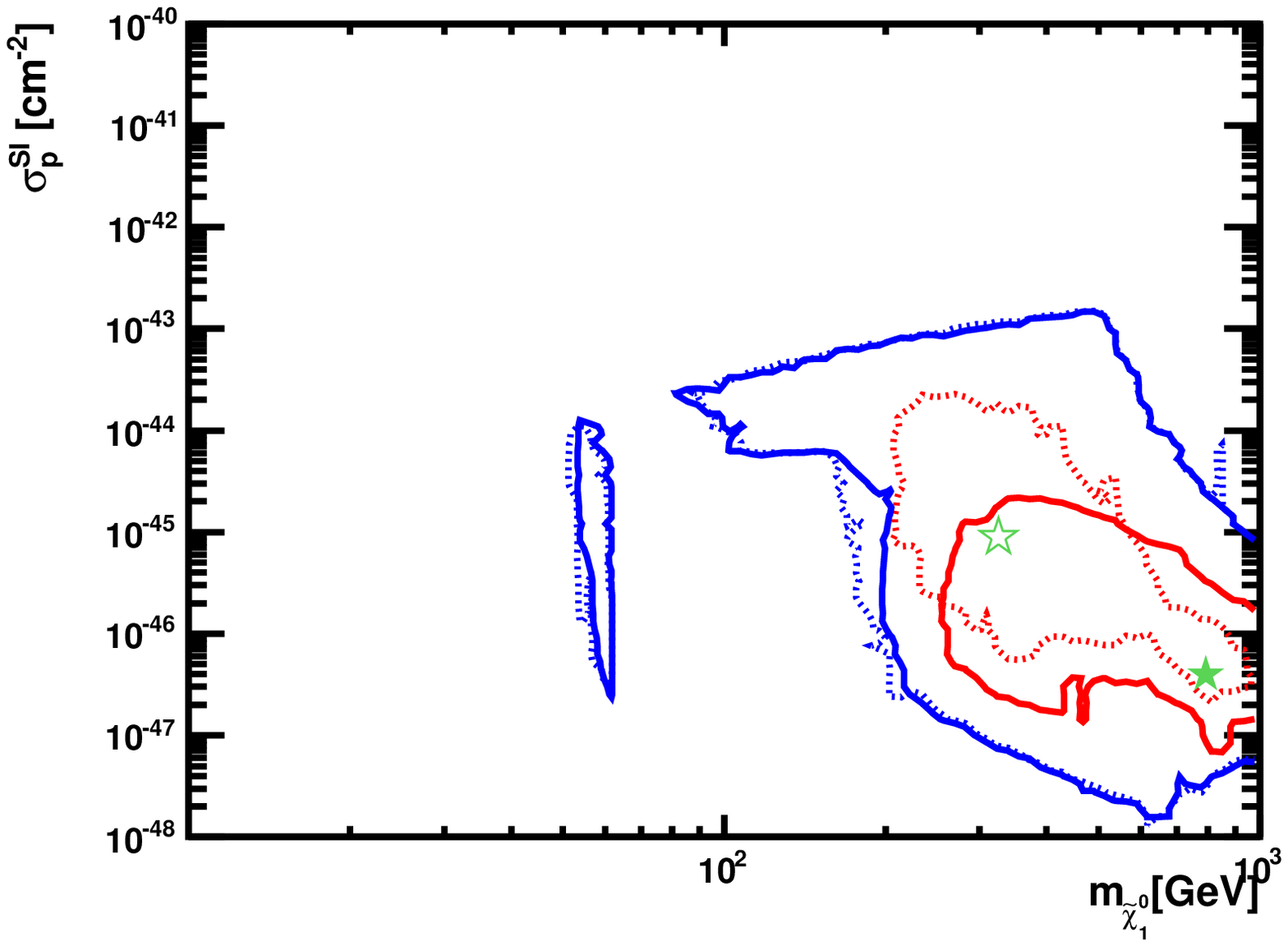}}
\resizebox{8.6cm}{!}{\includegraphics{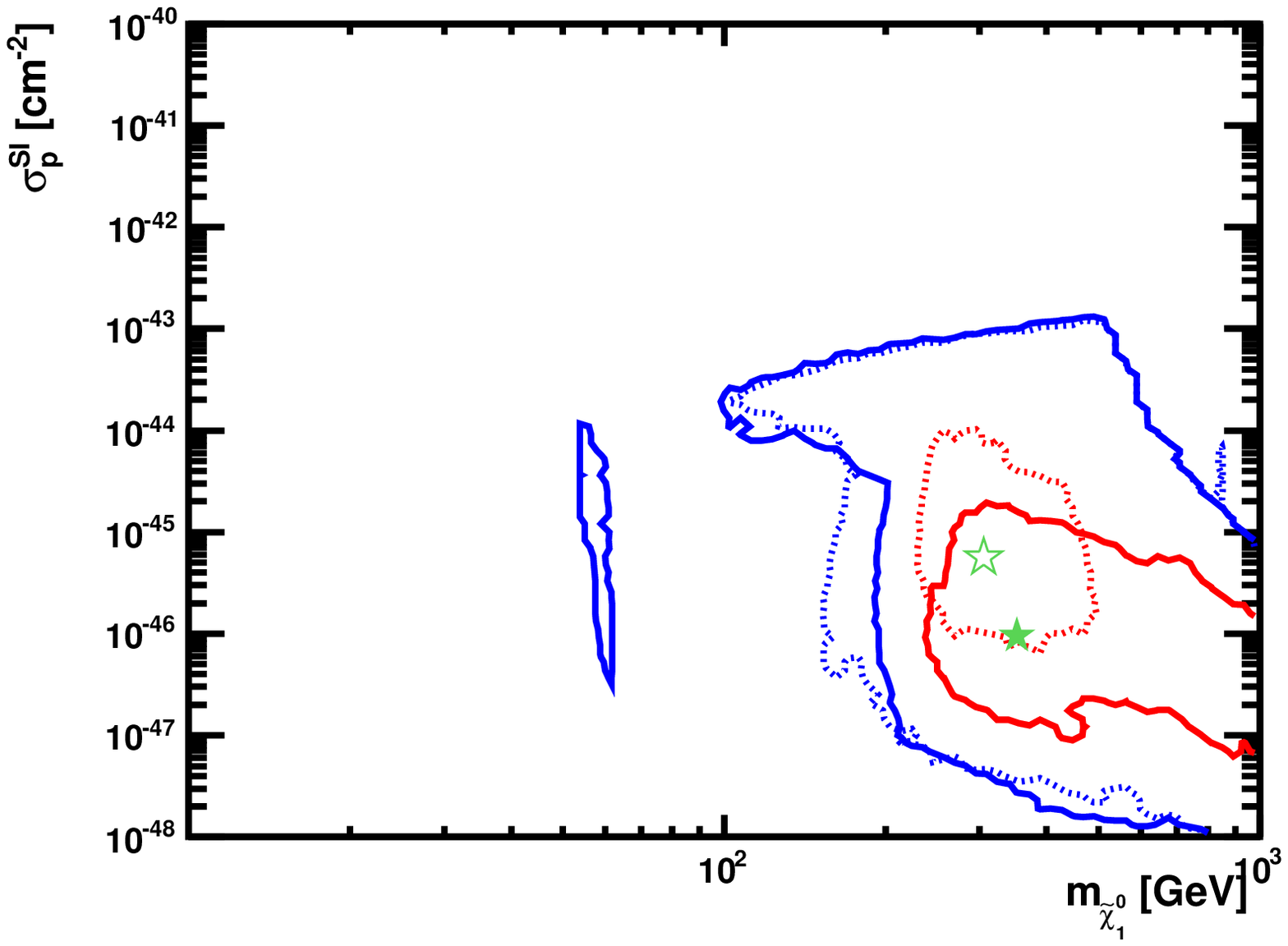}}
\vspace{-1cm}
\caption{\it The $(\mneu{1}, \ssi)$ planes in the CMSSM (left) and the
  NUHM1 (right), for $\Mh \simeq 125 \gev$. 
  The notations and significations of the contours are
  the same as in Fig.~\protect\ref{fig:ssi} \cite{mc7.5}. 
}
\label{fig:mneussi}
\end{figure*}

In Fig.~\ref{fig:mneussi} we show results for the
preferred regions in the $(\mneu{1}, \ssi)$ plane. 
As seen in Fig.~\ref{fig:mneussi}, the fact that larger values of $m_{1/2}$
and hence $\mneu{1}$ are favoured by the larger values of $\Mh$ implies that
at the 68\% CL the preferred range of $\ssi$\ is significantly lower when
$\Mh \simeq 125 \gev$, when compared to our previous best fit
with $\Mh = 119 \gev$,  rendering direct detection of dark matter
significantly more difficult.
Again, this effect on $\mneu{1}$ is more pronounced in the CMSSM, whereas in the NUHM1 
the value of $\mneu{1}$ for the best-fit point changes only
  slightly.

\begin{figure*}[htb!]
\resizebox{8.6cm}{!}{\includegraphics{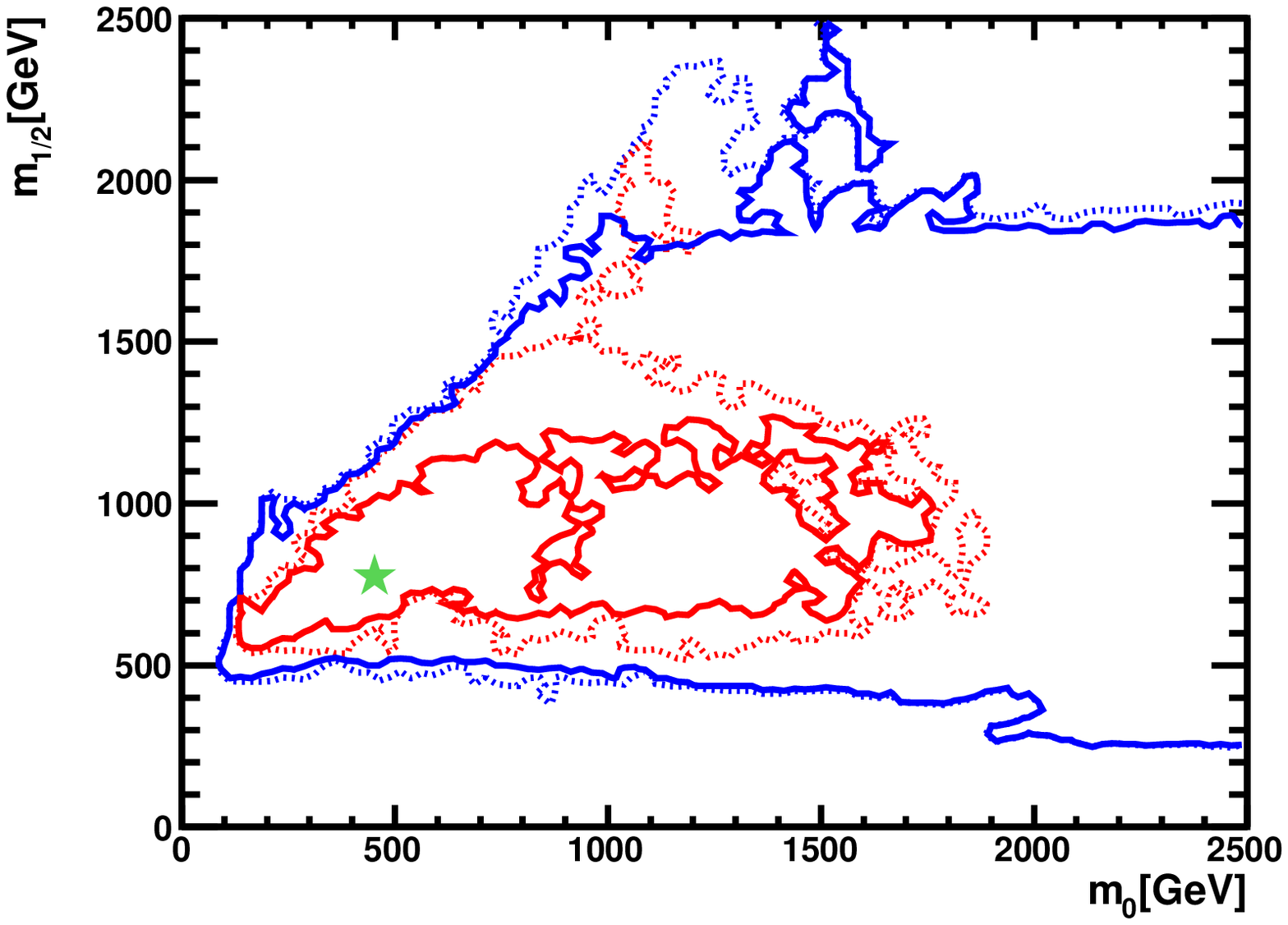}}
\resizebox{8.6cm}{!}{\includegraphics{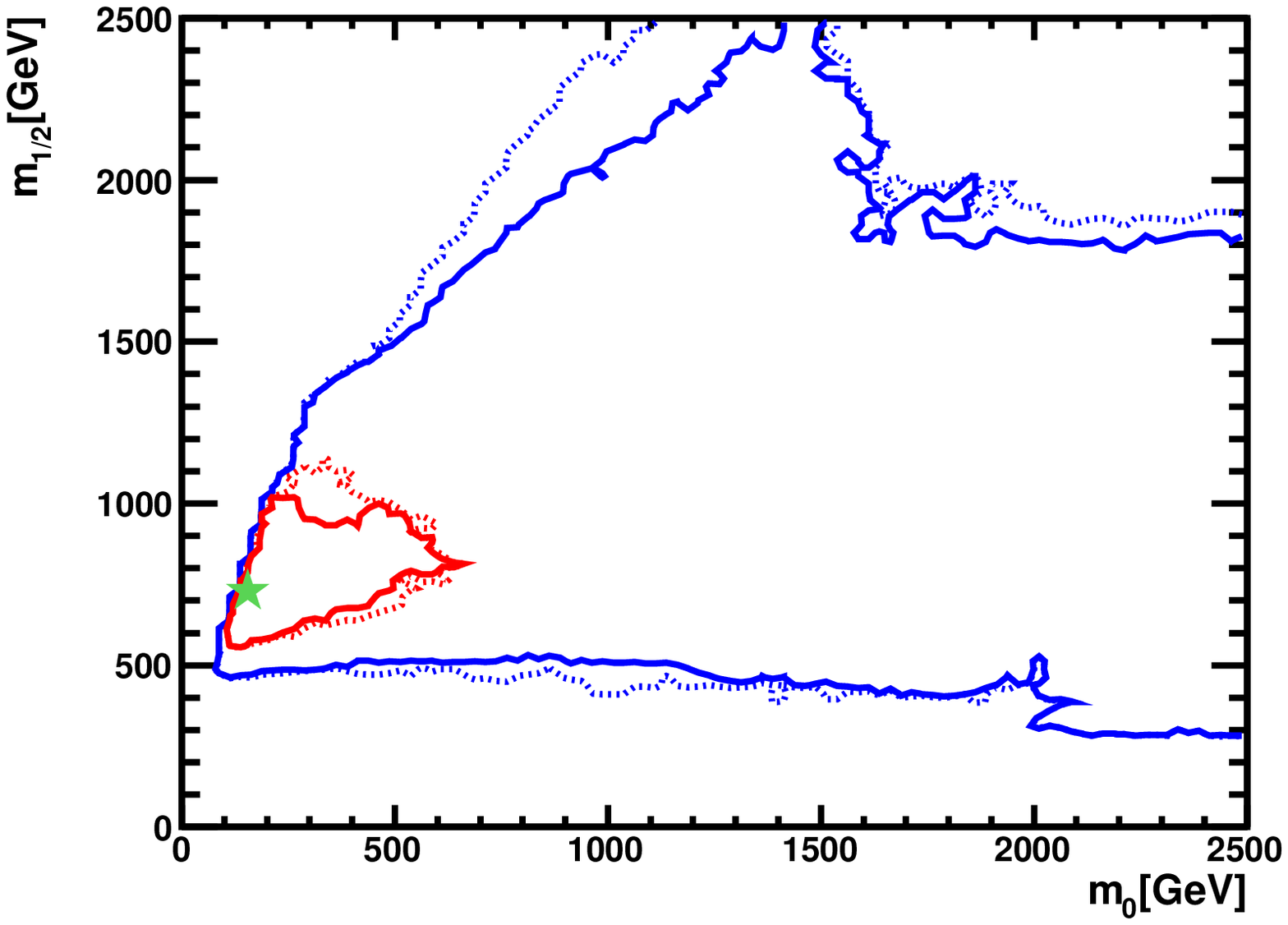}}
\caption{\it The $(m_0, m_{1/2})$ planes in the CMSSM (left) and the
  NUHM1 (right) assuming a hypothetical measurement of $\Mh = 119 \pm 1 \gev$.   
  The notations and significations of the contours are
  the same as in Fig.~\protect\ref{fig:6895} \cite{mc7.5}. 
}
\label{fig:6895119}
\end{figure*}

If instead, we assume an alternative potential
LHC measurement $\Mh = 119 \pm 1 \gev$, which corresponds to the CMS
$ZZ^*$ signal and our earlier predictions including the \gmt\ constraint, we obtain the 
$(m_0, m_{1/2})$ planes
shown in Fig.~\ref{fig:6895119}.
Since this assumed LHC
value of $\Mh$ coincides with the previous best-fit values in both
the CMSSM and NUHM1, the best-fit points in these models
(indicated by the green stars in Fig.~\ref{fig:6895119}) are
unaffected by the imposition of the putative LHC constraint.~%
The effect of the hypothetical measurement restricting the range 
in $m_{1/2}$ is indeed seen in both panels of Fig.~\ref{fig:6895119}, 
though
for the 68\% CL contour (shown in red) it is much more pronounced for
the CMSSM than for the NUHM1.

\begin{figure*}[htb!]
\resizebox{8.6cm}{!}{\includegraphics{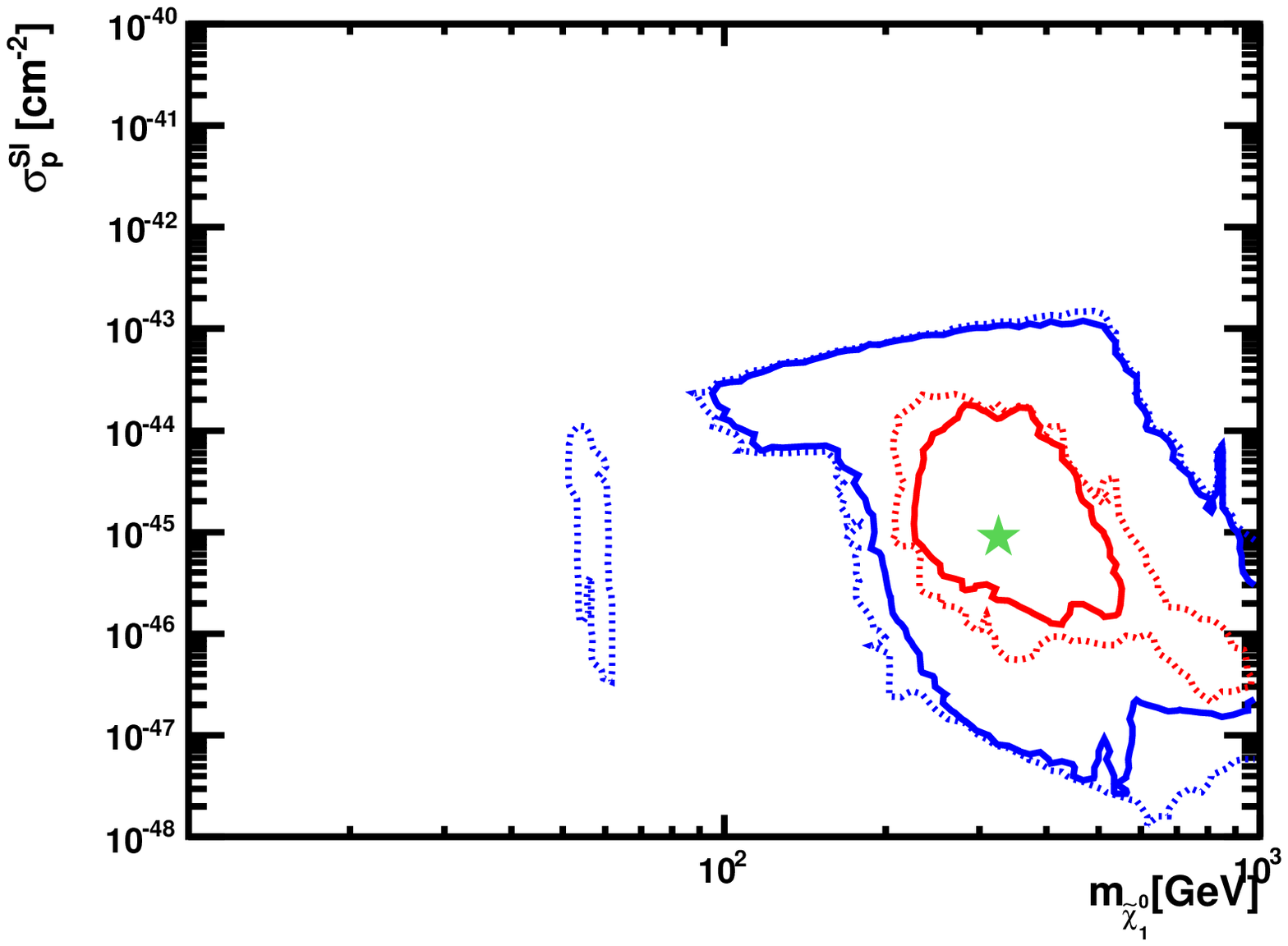}}
\resizebox{8.6cm}{!}{\includegraphics{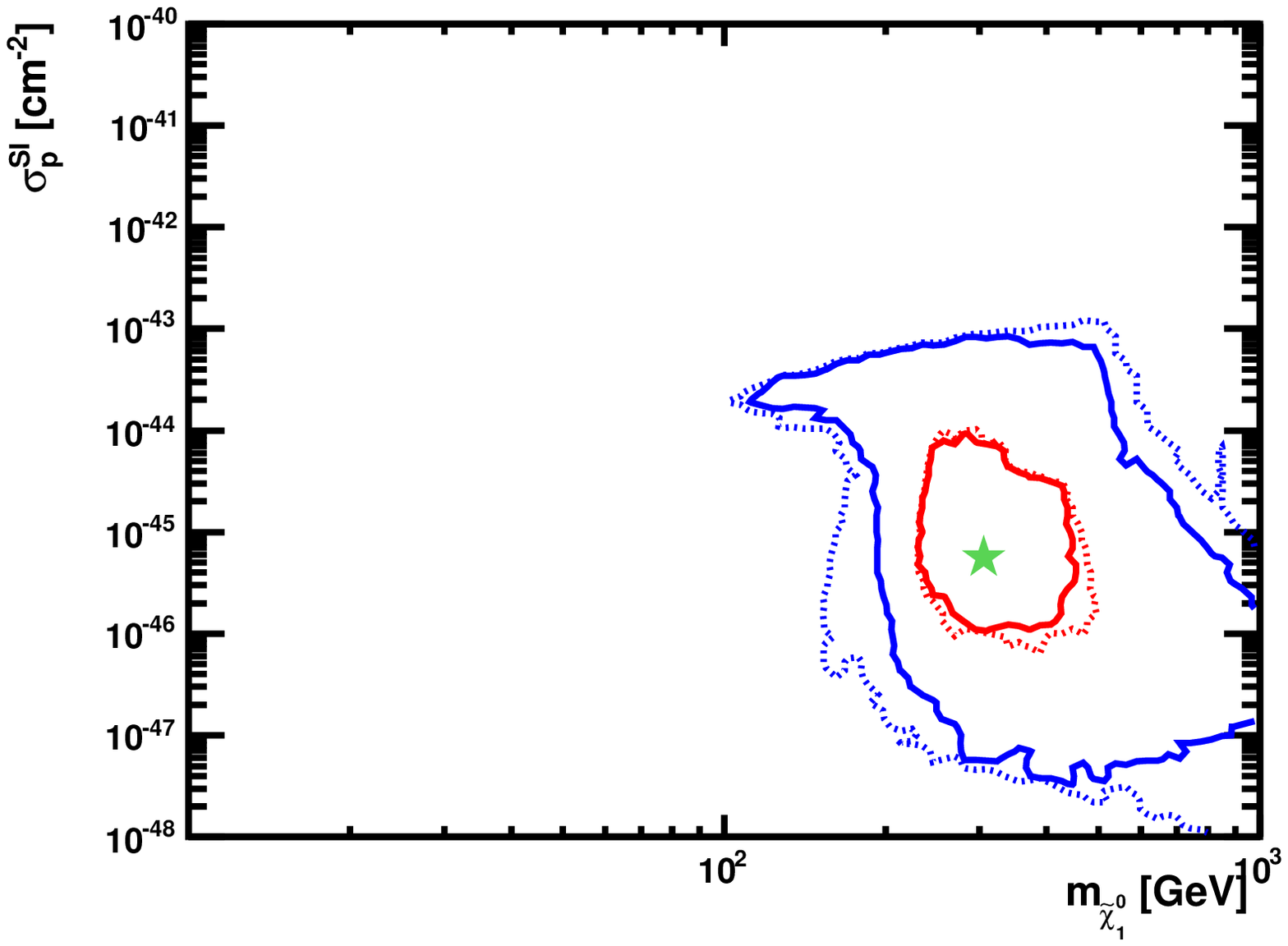}}
\vspace{-1cm}
\caption{\it The $(\mneu{1}, \ssi)$ planes in the CMSSM (left) and the
  NUHM1 (right), for $\Mh \simeq 119  \gev$. 
  The notations and significations of the contours are
  the same as in Fig.~\protect\ref{fig:6895119} \cite{mc7.5}. 
}
\label{fig:mneussi119}
\end{figure*}

In this case, 
in both the CMSSM
and the NUHM1 there is little impact on the 95\% CL regions  
nor on the 68\% CL region in the NUHM1 in the ($m_\chi,\ssi)$ plane. 
The only substantial change, as can be seen in
  Fig.~\ref{fig:mneussi119}, appears in the 68\% CL region of the CMSSM,
where now values of $\mneu{1} \ga 500 \gev$ and
$\ssi \lsim 10^{-46} {\rm cm}^{-2}$ are disfavoured after the inclusion of a
Higgs-boson mass measurement at $119 \gev$.

The past year has shown an immense amount of activity.
We have seen direct detection experiments (XENON100 \cite{Xenon100new})
for the first time have a direct impact on supersymmetric parameter space making 
these data a necessary input to a global likelihood analysis.
The LHC constraints have moved at a frightening pace.
As reviewed here, starting with the 35/pb data, our notion of the
best fit point in the CMSSM and indeed our prospects for low energy
supersymmetry are greatly different as we start 2012 compared
with the pre-LHC era.

\ack I would like to thank all the members of Mastercode collaboration whose
work went into the summary presented here.
This work was supported in part
by DOE grant DE--FG02--94ER--40823 at the University of Minnesota.

\end{document}